\documentclass[final,5p,twocolumn,sort&compress]{elsarticle}

\usepackage{leftindex}
\usepackage{scalerel}
\usepackage{tabularx}
\usepackage{supertabular}
\usepackage[math]{cellspace}

\usepackage{xspace}
\usepackage[dvipsnames]{xcolor}
\usepackage{hyperref}

\hypersetup{
    colorlinks=true, 
    linkcolor=royalblue, 
    citecolor=magenta}

\newcommand{\LCDM}{$\Lambda$CDM\xspace}

\journal{Physics of the Dark Universe}

\begin{document}

\begin{frontmatter}

\title{The interacting vacuum and  tensions: a comparison of theoretical models}

\author[Sussex]{Marco Sebastianutti \corref{cor}}
\author[LUPM]{Natalie B. Hogg}
\author[ICG,INFN]{Marco Bruni}

\address[Sussex]{Department of Physics \& Astronomy, University of Sussex, Brighton, BN1 9QH, United Kingdom}

\address[LUPM]{Laboratoire Univers et Particules de Montpellier, Universit\'{e} de Montpellier and CNRS, Montpellier, France}

\address[ICG]{Institute  of  Cosmology  and  Gravitation,  University  of  Portsmouth, Burnaby  Road,  Portsmouth,  PO1  3FX, United Kingdom}

\address[INFN]{INFN Sezione di Trieste, Via Valerio 2, 34127 Trieste, Italy}

\cortext[cor]{Corresponding author: m.sebastianutti@sussex.ac.uk}







\begin{abstract}
    We analyse three interacting vacuum dark energy models with the aim of exploring whether the $H_0$ and $\sigma_8$ tensions can be simultaneously resolved in such models. We present the first ever derivation of the covariant gauge-invariant perturbation formalism for the interacting vacuum scenario, and, for the sub-class of geodesic cold dark matter models, connect the evolution of perturbation variables in this approach to the familiar cosmological observables. We show how $H_0$ and $\sigma_8$ evolve in three interacting vacuum models: firstly, a simple linear coupling between the vacuum and cold dark matter; secondly, a coupling which mimics the behaviour of a Chaplygin gas; and finally a coupling which mimics the Shan--Chen fluid dark energy model. We identify, if any, the regions of parameter space which would correspond to a simultaneous resolution of both tensions in these models. When constraints from observational data are added, we show how all the models described are constrained to be close to their $\Lambda$CDM limits.
\end{abstract}



\begin{keyword}
    dark energy theory \sep cosmological perturbation theory in GR and beyond
\end{keyword}

\end{frontmatter}


\section{Introduction}
\label{sec:intro}

The \LCDM model, based on inhomogeneities growing in an otherwise spatially homogeneous and isotropic Friedmann--Lema\^itre--Robertson--Walker (FLRW) background spacetime, explains two of the main features which are observed in our Universe: an acceleration of the expansion rate at late times, and the formation of large-scale structures.

The late-time accelerated expansion of the Universe, discovered by the realisation that the luminosity of Type Ia supernovae (SNIa) standardisable candles was dimmer than expected~\citep{Riess1998,Perlmutter1999}, and further confirmed by observations of the cosmic microwave background (CMB)~\citep{Dodelson:1999am} and baryon acoustic oscillations (BAO)~\citep{Eisenstein:2005su}, is explained within the framework of General Relativity (GR) by means of a dark energy component that violates the strong energy condition, see e.g.~\cite{Novello:2008ra}.

In practice, for a dark energy  described by the perfect fluid energy--momentum tensor, as later described in equation \eqref{eq:perf_fluid}, this condition implies that its  equation of state parameter $w\equiv p/\rho$, where $p$ is the pressure and $\rho$ the energy density of the fluid, must be $w\leq -1/3$. The simplest way to satisfy this requirement is by invoking the presence of a constant vacuum energy density $V=c^4\Lambda/8\pi G_{\rm N}$ characterised by $w=-1$, where $c$ is the speed of light and $G_{\rm N}$ the Newton's constant. In $\Lambda$CDM, this constant vacuum energy density is the cosmological constant $\Lambda$, 
yet whether this component is the correct explanation for the late-time accelerated expansion of the Universe is yet to be understood.


Furthermore, observations of large-scale structure, along with galaxy evolution and strong gravitational lensing, have provided robust evidence for the existence of the other major unknown component in the Universe: cold dark matter (CDM)~\citep{Zwicky,Eke1998,Sofue:2000jx,Frenk:2012ph,Kunz:2016yqy, Massey:2010hh}. In the standard picture, CDM is described by a weakly interacting massive particle with an equation of state $w=0$, and occupies a further $25\%$ of the total matter--energy budget of the Universe.

Despite its successes in explaining observational phenomena, the \LCDM model faces a number of challenges: the discrepancy between the predicted and observed values of the cosmological constant, also known as the cosmological constant problem \citep{Weinberg:1988cp}; and  the tensions that exist between low-redshift probes of the expansion rate and structure growth and the corresponding values inferred from CMB measurements -- the $H_0$ and $\sigma_8$ tensions \citep{Macaulay:2013swa,Riess:2016jrr,Bernal:2016gxb} (see e.g. \cite{Abdalla:2022yfr} for a review).

With current observational data, these tensions now stand at around $5\sigma$ in $H_0$, where the discrepancy is between measurements from the CMB (made in the framework of \LCDM) \citep{Planck:2018vyg} and the higher values coming from Cepheid-calibrated SNIa \citep{Riess:2021jrx}; and at around $2\sigma$ in $\sigma_8$, where the discrepancy lies between CMB measurements\footnote{A recent analysis of the Planck data using a new likelihood found the $\sigma_8$ tension to be decreased to around $1.6\sigma$~\citep{Tristram:2023haj}, while CMB lensing measurements from ACT, which probes low redshift structure, found no evidence for a tension at all \citep{Qu2023}.} \citep{Planck:2018vyg} and the lower values coming from various galaxy clustering and weak lensing measurements made by KiDS \citep{Heymans:2020gsg} and DES \citep{DES:2020ahh}.

Motivated by these problems with \LCDM cosmology, many alternative models have been proposed to explain the late-time accelerated expansion of the Universe, often with the further aim of resolving or relaxing the cosmological tensions. A common feature of these models is the addition of a field or fields whose evolution modifies the dark energy equation of state, resulting in extra degrees of freedom. Some well-known examples of dynamical dark energy models driven by an additional scalar field are quintessence~\citep{Amendola:1999er} and k-essence~\citep{Armendariz-Picon:2000ulo}; see~\cite{Copeland:2006wr} for a review.

A different approach is to consider dark energy and CDM as a single cosmological fluid, such as a unified dark matter which behaves according to a given equation of state. This idea dates back to the introduction of the Chaplygin gas model \citep{Kamenshchik:2001cp,Bento:2002ps} in which the transition from matter to dark energy domination is achieved using a perfect fluid with a non-standard equation of state. Another example of a cosmological matter--energy fluid invoked to explain the late-time accelerated expansion comes from a non-linear equation of state borrowed from the field of lattice kinetic theory~\citep{Shan1993}, and is known as the Shan--Chen dark energy model \citep{Bini:2013ods,Bini:2016wqr}.

Besides the use of a single fluid to explain the behaviour of both CDM and dark energy, these two components can be coupled, or interact, via an exchange of energy, momentum or both. One class of interacting dark energy models is the so-called interacting vacuum scenario \citep{Wands:2012vg}. An attractive advantage of this class of models is that there are no additional dynamical degrees of freedom compared to a cosmological constant, hence the cosmology reduces to \LCDM when the interaction vanishes, i.e. when $V = \Lambda/\kappa=$ constant, where $\kappa=8\pi G_{\rm N}/c^4$.

The aim of this work is to compare and contrast the cosmological dynamics of three different interacting dark energy models. We aim to demonstrate why the cosmological tensions may remain unresolved in such models, or point out areas of parameter space which may lead to tension resolution. To this end, we provide a covariant gauge-invariant description of the interacting vacuum and, by assuming CDM remains geodesic and comoving with the baryonic component, study the evolution of linear matter density perturbations in this framework. The models we choose to study are: a simple one with interaction linear in the vacuum energy density, which we dub the interacting linear vacuum model 
\citep{Salvatelli:2014zta,Martinelli:2019dau,Kaeonikhom:2022ahf}, the interacting generalised Chaplygin gas model \citep{Bento:2004uh,Wang:2013qy,Wang:2014xca}, and the interacting Shan--Chen model \citep{Hogg:2021yiz}.

This work is organised as follows: in section \ref{sec:IV} we present the covariant and gauge-invariant description of the interacting vacuum, deriving the background and linear perturbations in this approach initially leaving the interaction unspecified and later specialising to the geodesic CDM case. In section \ref{sec:method} we describe the interacting dark energy models studied in this work and outline the method we use to analyse them. In section \ref{sec:results} we present the results of our analysis, focusing on the evolution of quantities such as $h$ and $f\sigma_8$ as functions of scale factor and coupling strengths in these models. In section \ref{sec:Conlusions} we discuss our results and make some concluding remarks. Following~\cite{Bruni:1992dg,Dunsby:1991xk}, in \ref{app:A}, we demonstrate the equivalence of the covariant gauge-invariant (CGI) and Bardeen variables at first order; in \ref{app:B} we provide a complete derivation of the first order ODEs for the CGI density and expansion perturbations $\Delta_\mu$ and $Z_\mu$; in \ref{app:C} we list the identities, valid at linear order in perturbations and for CGI quantities, used to obtain the results presented in the main text, and in \ref{app:D} we show that a geodesic interaction automatically results in a synchronous gauge.

\paragraph{Conventions} Throughout this work, we adopt the signature $(-,+,+,+)$ for the
spacetime metric.
In section \ref{sec:IV} of this paper we use units in which $c=1$; from section \ref{sec:method} onwards we adopt a geometrised unit system where both $c$ and $8\pi G_{\rm N}$ are set to unity. For cosmological parameters such as $H_0$, we adopt the subscript ‘0’ to denote observables evaluated today, i.e. at redshift $z=0$ or equivalently scale factor $a=1$. Finally, in this paper we neglect the radiation component, i.e. we set $\Omega_{\rm r, 0}=0$.

\section{Interacting vacuum and its covariant and gauge-invariant perturbations}
\label{sec:IV}
Perturbation theory is an essential part of cosmology, allowing us to understand the formation and growth of the observed structure on the largest scales. The common way to deal with perturbations is to specify a background spacetime and decompose physical quantities into a homogeneous background and inhomogeneous perturbations. Although GR is covariant, i.e. manifestly independent from the coordinate choice, splitting variables into a background part and a perturbation is not a covariant procedure, and therefore introduces a coordinate or gauge dependence, as first geometrically elucidated by \cite{R.K.Sachs_1964}. Observable geometrical and physical quantities should then be described by gauge-invariant variables \citep{S.W.Hawking_1966,S.A.Teukolsky_1973,Stewart:1974uz,Bardeen:1980kt,F.V.Mukhanov_etal_1992,R.M.Wald_1984,Ellis:1989jt,J.M.Stewart_1990,M.Bruni_etal_1997,M.Bruni_S.Sonego_1999,Malik:2008tq,ellis_maartens_maccallum_2012,J.Yoo_R.Durrer_2017}.

In order to deal with this problem, Bardeen-type variables can be used \citep{Bardeen:1980kt}. These are linear combinations of gauge-dependent quantities specifically constructed to be gauge-invariant variables {\it at first order}. Although this procedure can be extended to higher order \citep{Malik:2008tq}, the use of the Bardeen formalism can make it difficult to interpret the results of a given calculation, since the physical and geometrical meaning of the variables is often obscure unless a gauge is specified. 

An alternative approach is  to use covariantly-defined spatial gradients of  physical quantities, such as the energy density, the expansion scalar or the 3-Ricci curvature. In this approach no approximation is used in defining the variables, which are then exactly defined in any spacetime.\footnote{In this sense, this approach is similar to that of \cite{S.A.Teukolsky_1973} and \cite{Stewart:1974uz} for black hole perturbations, see \cite{P.Pani_2013} for a review.} This means that no non-covariant splitting is performed, and perturbations can be computed covariantly, allowing the formalism to be specified to any background. This method is known as the covariant gauge-invariant (CGI) approach \citep{Ellis:1989jt,Bruni:1992dg,Bruni:1991kb,Dunsby:1991xk,ellis_maartens_maccallum_2012}, extending and completing Hawking's pioneering approach \citep{S.W.Hawking_1966}; see \cite{D.Langlois_F.Vernizzi_2010} for a review.

In this section, we develop the specific covariant and gauge-invariant perturbations formalism for the interacting vacuum scenario, first reviewing the covariant construction of a vacuum component interacting with a perfect fluid \citep{Wands:2012vg}. Specialising to a spatially flat FLRW background, we further restrict our study to that of the geodesic CDM case by considering a pure energy interaction.

\subsection{Covariant description of the interacting vacuum}

Following \cite{Wands:2012vg}, let us now consider a covariant description of both the perfect fluid and vacuum components, as well as the interaction between the two.

The energy--momentum tensor of a perfect fluid is: 
\begin{equation}
    T_{\mu\nu}=(\rho+p)u_\mu u_\nu+p\,g_{\mu\nu}, \label{eq:perf_fluid}
\end{equation}
where $\rho$ is the fluid energy density and $p$ its pressure, which vanishes in the case of a CDM (or baryonic) component;  $u^{\mu}$ is the 4-velocity of the fluid and the eigenvector of $T_{\mu\nu}$ which defines the fluid rest frame.
The energy--momentum tensor of a vacuum component, $\check{T}_{\mu \nu}$, is proportional to the metric
\begin{equation}
\label{V}
    \check{T}_{\mu\nu}=-V g_{\mu\nu},
\end{equation}
and by comparison with the perfect fluid energy--momentum tensor in \eqref{eq:perf_fluid} we can identify the vacuum energy density and pressure as $\rho = V = -p$; its equation of state is thus $w\equiv p/\rho=-1$. 

The interaction between the perfect fluid and the vacuum energy is introduced in the following way:
\begin{align}
    \nabla_\mu T^{\mu}_{\,\,\,\nu}&=-Q_\nu,
    \label{int1}
    \\
    \nabla_\mu \check{T}^{\mu}_{\,\,\,\nu}&=-\nabla_\nu V=Q_\nu,
    \label{int2}
\end{align}
where the interaction 4-vector $Q^\mu$ stands for the energy--momentum flow between the two components.
This scenario reduces to the standard \LCDM non-interacting limit when the interaction vanishes, i.e.\ $Q^\mu=0$ and $V=\Lambda/\kappa=$ constant. Adding \eqref{int1} to \eqref{int2} we immediately recover the covariant energy--momentum conservation relation
\begin{equation}\label{tot}
    \nabla_\mu T^{\mu\nu}_{\,\,\text{(tot)}}=\nabla_\mu( T^{\mu\nu}+\check{T}^{\mu\nu})=0,
\end{equation}
as implied by the Einstein equations and the Bianchi identities.

In general, for any given 4-velocity $n^\mu$ (a timelike vector), one can project a tensorial quantity into a part parallel to $n^\mu$ and a part orthogonal to it by the use of a projector tensor $P^\mu_{\,\,\,\nu}\equiv \delta^\mu_{\,\,\,\nu} +n^\mu n_\nu$. Using specifically the fluid 4-velocity $u^\mu$, we can then define the projector
\begin{equation}\label{eq:hmunu}
 h_{\mu\nu}\equiv  g_{\mu\nu}+u_\mu u_\nu,
\end{equation}
and then the energy--momentum exchange $Q^\mu$ can be decomposed into two parts,
\begin{equation}\label{Q}
    Q^\mu=Qu^\mu+f^\mu.
\end{equation}
For observers comoving with the perfect fluid, $Q=-Q_\mu u^\mu$ represents the energy exchange between the two components and $f^\mu=h^\mu_{\,\,\,\nu}Q^\nu$ represents the momentum transfer; from the above definition we have $f_\mu u^\mu=0$. In this paper we shall focus on the case of a pure energy exchange in the frame of the fluid $u^\mu$, $f^\mu=0$.

It is clear from \eqref{V} that any timelike 4-vector is an eigenvector of $\check{T}_{\mu\nu}$ with eigenvalue $V$. This means that $V$ is the energy density of vacuum for any observer; in other words, the vacuum energy does not have a unique 4-velocity, or else, absolute motion with respect to vacuum could be detected, as observed by \cite{Lemaitre1933}. However, one can use $Q^\mu=-\nabla^\mu V$ to define a preferred 4-velocity for the vacuum, 
\begin{equation}
\label{vacuumvelocity}
    \check{u}^\mu=\frac{-\nabla^\mu V}{\sqrt{\lvert \nabla_\nu V\nabla^\nu V \rvert}},
\end{equation}
so that in the $\check{u}^\mu$ frame there is no momentum transfer, as shown in Eq.\ \eqref{CGIvacuum2}.

\subsection{Covariant gauge-invariant formalism}
\label{sec:CGI}
Having defined the covariant vacuum, we can proceed to the CGI formalism for perturbations. Central to the CGI approach is the  Stewart and Walker lemma \citep{Stewart:1974uz} for the first order perturbation $\delta\textbf{T}$ of a tensorial quantity $\textbf{T}$ about a given background spacetime, where $\textbf{T}$ acquires value $\textbf{T}_0$;
the lemma states that: 
\begin{quote}
    The linear perturbation $\delta \textbf{T}$ of a tensorial quantity $\textbf{T}$ with value $\textbf{T}_0$ on the background spacetime is gauge-invariant if and only if one of the following holds:
    \begin{enumerate}
        \item $\textbf{T}_0$ vanishes;
        \item $\textbf{T}_0$ is a constant scalar; or
        \item $\textbf{T}_0$ is a constant linear combination of Kronecker deltas.
    \end{enumerate}
\end{quote}
This simply follows from the fact that if $\boldsymbol{\xi}$ is the generator of the gauge transformation $\Tilde{x}^\mu=x^\mu+\lambda \xi^\mu$ on the background spacetime, where $\lambda$ is an expansion parameter, then in general $\delta\Tilde{\textbf{T}}=\delta \textbf{T}+\mathcal{L}_{\boldsymbol{\xi}} \textbf{T}_0$, where $\mathcal{L}_{\boldsymbol{\xi}} \textbf{T}_0$ is the Lie derivative of $\textbf{T}_0$ along $\boldsymbol{\xi}$, see \cite{M.Bruni_etal_1997,Matarrese:1997ay, M.Bruni_etal_2003_17Jan,C.F.Sopuerta_etal_2003} for more details and the extension to higher orders.

In the following, after introducing the covariant spacetime splitting, we will specialise to the FLRW background. In order to define CGI perturbations of the latter, according to the first condition of the lemma we simply need to find covariantly defined quantities which vanish in FLRW.\footnote{We do not consider the second and third conditions of the lemma, as in an FLRW Universe the only invariantly defined constant is the cosmological constant and no tensors that are constant products of Kronecker deltas occur naturally.} In this formalism, therefore, one starts from exact nonlinear equations for these quantities, then, the linearisation scheme simply consists of linearising these equations with respect to the CGI variables. Quantities which do not vanish in the FLRW background spacetime, such as the  fluid density $\rho$ and pressure $p$, are instead retained in the equations at zeroth order only.

\subsubsection{Covariant variables}
The CGI approach is based on a covariant (1+3) spacetime splitting which is realised by projecting tensorial quantities both along the preferred perfect fluid 4-velocity $u^\mu$ and, using the projection tensor \eqref{eq:hmunu}, on the spatial local subspace orthogonal to $u^\mu$. 
Adopting this approach \citep{S.W.Hawking_1966,G.F.R.Ellis_2009,Ellis:1989jt,D.Langlois_F.Vernizzi_2010,ellis_maartens_maccallum_2012}, we can split the 4-velocity covariant derivative, $\nabla_\nu u_\mu$, into two parts: one giving the variation along the time direction defined by $u^\mu$ and, by means of the spatial gradient $\leftindex^{(3)}\nabla_\nu\equiv h_{\nu}^{\,\,\,\mu}\nabla_\mu$, one containing spatial variations only:
\begin{align}\label{u-split}
\nabla_{\nu}u_\mu&=\leftindex^{(3)}\nabla_\nu u_\mu-a_{\mu}u_{\nu},
\end{align}
$a^\nu\equiv u^\mu\nabla_\mu u^\nu$ being the 4-acceleration of the fluid. The spatial part is then split into a symmetric and an anti-symmetric part:
\begin{align}\label{spatial-u-split}
    \leftindex^{(3)}\nabla_\nu u_\mu&=\Theta_{\mu\nu}+\omega_{\mu\nu},
\end{align}
where $\Theta_{\mu\nu}=h_{\mu}^{\,\,\,\rho}h_{\nu}^{\,\,\,\sigma}\nabla_{(\rho}u_{\sigma)}$, and $\omega_{\mu\nu}=h_{\mu}^{\,\,\,\rho}h_{\nu}^{\,\,\,\sigma}\nabla_{[\rho}u_{\sigma]}$ is the vorticity of the fluid. If $\omega_{\mu\nu}=0$, then $u^\mu$ is hypersurface orthogonal and $h_{\mu\nu}$ is the 3-metric on the hypersurfaces orthogonal to $u^\mu$, see \cite{R.M.Wald_1984,ellis_maartens_maccallum_2012}.
We can further decompose the symmetric part into a symmetric traceless and a trace part, the shear $\sigma_{\mu\nu}$ and the expansion scalar $\Theta$ respectively,
\begin{align}
\Theta_{\mu\nu}&=\sigma_{\mu\nu}+\frac{1}{3}\Theta h_{\mu\nu}.
\end{align}
Putting this all together yields:
\begin{align}
    \nabla_{\nu}u_\mu&=\omega_{\mu\nu}+\sigma_{\mu\nu}+\frac{1}{3}\Theta h_{\mu\nu}-a_{\mu}u_{\nu}.
\end{align}
From the Stewart and Walker lemma, the exactly defined kinematical quantities $\omega_{\mu\nu}$, $\sigma_{\mu\nu}$ and $a_\nu$ are all CGI  variables at first order.
Other relevant CGI quantities are defined using the spatial gradient operator $\leftindex^{(3)}\nabla_\mu$:
\begin{align}\label{list}
    Z_\mu &\equiv a\leftindex^{(3)}\nabla_\mu\,\Theta, \nonumber \\ 
    R_\mu &\equiv a^3\leftindex^{(3)}\nabla_\mu^{\,\,\,\,(3)}R, \nonumber \\
    P_\mu &\equiv \frac{a}{p}\leftindex^{(3)}\nabla_\mu\,p, \nonumber \\
    \Delta_\mu &\equiv \frac{a}{\rho}\leftindex^{(3)}\nabla_\mu\,\rho,
\end{align}
which are the comoving expansion, 3-curvature, fractional pressure and density spatial gradients respectively.
The latter is the comoving CGI analogue of the density contrast $\delta=\delta\rho/\rho$ at first order; it is directly related to Bardeen's gauge-invariant density perturbation $\Delta_{\textcolor{black}{\rm B}}$ ($\epsilon_{\rm m}$ in the language of~\cite{Bardeen:1980kt}). See \ref{app:A} for the derivation and \cite{Bruni:1992dg} for the complete list of equivalencies between the CGI and Bardeen approaches.

Similarly, we can define the vacuum energy spatial gradient
\begin{align}\label{CGIvacuum1}
    V_\mu \equiv \leftindex^{(3)}\nabla_\mu\,V
    =-h^{\,\,\,\nu}_{\mu}\,Q_\nu=-f_\mu,
\end{align}
where the last two equalities follow from \eqref{int2} and \eqref{Q}.
The above expression shows that vacuum inhomogeneities in the fluid frame arise due to the non-vanishing momentum transfer $f_\mu$. If, instead of the $h_{\mu\nu}$ projector, we consider $\check{h}_{\mu\nu}$, defined to be orthogonal to $\check{u}^\mu$, from \eqref{vacuumvelocity} we obtain
\begin{align}\label{CGIvacuum2}
    \check{V}_\mu\equiv\check{h}^{\,\,\,\nu}_{\mu}\,\nabla_\nu V=-\check{h}^{\,\,\,\nu}_{\mu}\,\check{u}_\nu\,\lvert \nabla_\rho V\nabla^\rho V \rvert^{1/2}=0,
\end{align}
which vanishes by construction. Therefore in the $\check{u}^\mu$ frame of reference the vacuum energy density remains unperturbed at all orders in perturbations, not just linearly.

\subsubsection{Fundamental equations}
Projecting the energy--momentum conservation relations \eqref{int1} and \eqref{int2} along $u^\mu$, and \eqref{tot} with $h_{\mu\nu}$, we obtain
\begin{align}
    \dot{V}&=Q,\label{continuity1}\\
    \dot{\rho}&=-\Theta\left(\rho+p\right)-Q,\label{continuity2}\\
    a_\mu&=\left(\rho+p\right)^{-1}\left(V_\mu-Y_\mu\right),\label{acceleration}
\end{align}
which are the vacuum and perfect fluid continuity equations, and a dynamical relation for the previously defined 4-acceleration respectively with $Y^\mu\equiv\leftindex^{(3)}\nabla^\mu\,p$; the dot represents the covariant derivative along $u^\mu$, e.g\ $\dot{\rho}=u^\mu\nabla_\mu \rho$, i.e. the derivative with  respect to proper time of the comoving observers. Together with \ref{CGIvacuum1}, \ref{acceleration} directly shows that pressureless matter -- in other words, dust -- remains geodesic when $f^\mu=0$, as in the non-interacting case.


The above evolution equations for $\rho$ and $V$ are coupled with the evolution equation for $\Theta$, given by the Raychaudhuri equation, which in the case of a perfect fluid coupled to the vacuum reads
\begin{align}\label{eq:Raychaudhuri}
    \dot{\Theta}+\frac{1}{3}\Theta^2+2\left(\sigma^2-\omega^2\right)-A+\frac{1}{2}\kappa\left(\rho+3p-2V\right)=0,
\end{align}
where $A\equiv\nabla_\nu\,a^\nu$ and $\sigma^2\equiv\frac{1}{2}\sigma_{\mu\nu}\sigma^{\mu\nu}$, $\omega^2\equiv\frac{1}{2}\omega_{\mu\nu}\omega^{\mu\nu}$ are the magnitudes of the shear and vorticity tensors respectively. Note that, assuming that the perfect fluid satisfies the strong energy condition, i.e. $(\rho+3p)\geq 0$, the contribution to the volume expansion due to  vacuum is opposite to the one coming from the perfect fluid.

The continuity and Raychaudhuri equations are in general also coupled to the equations for the shear and vorticity; the contributions of the CGI variables $\sigma_{\mu\nu}$ and $\omega_{\mu\nu}$ to \eqref{eq:Raychaudhuri} are however nonlinear, hence  they can be neglected at first order. Furthermore, $\omega_{\mu\nu}$ embodies vector perturbations, while we here focus on scalar perturbations only.

Finally, assuming $\omega_{\mu\nu}=0$, we can also write down the exact Hamiltonian constraint relating the spatial Ricci scalar of the hypersurfaces orthogonal to the perfect fluid flow $u^\mu$ with the shear and energy densities:
\begin{align}\label{eq:hamiltonian}
    ^{(3)}R=2\left[\sigma^2-\frac{1}{3}\Theta^2+\kappa\left(\rho+V\right)\right]. 
\end{align}
This shows that at first order, when we can neglect the CGI $\sigma^2$ term, there are only three degrees of freedom, e.g.\ the two energy densities and the 3-curvature, with the fourth variable given by the Hamiltonian constraint.

\subsubsection{Spatially flat FLRW background}
\label{sec:2.2.4}
We examine our results thus far in a spatially flat ($^{(3)}R=0$) FLRW background, in which shear and vorticity do not contribute and $a_\mu=0$. In such a background the expansion scalar is $\Theta=3H$ and the Raychaudhuri equation \eqref{eq:Raychaudhuri} becomes 
\begin{align}\label{eq:bgRay}
    \frac{\ddot{a}}{a}=\dot{H}+H^2=-\frac{\kappa}{6}(\rho-2V),
\end{align}
where $a$ is the cosmic scale factor, $H\equiv\dot{a}/a$ is the Hubble expansion rate and we consider a perfect fluid with vanishing pressure, $p=0$, which corresponds to non-relativistic matter or dust.

In both the non-interacting ($V=\Lambda/\kappa$) and the interacting ($V=V(t)$) case, the vacuum energy accelerates the expansion of spacetime, working against the action of matter. However, in the latter case, the acceleration can be enhanced or suppressed compared to the acceleration in the former case; this depends on the direction of the energy transfer between vacuum and matter and the specific form of the coupling function, as we will demonstrate below.

The Friedmann constraint equation in this FLRW background, obtained from \eqref{eq:hamiltonian}, simply reads 
\begin{align}\label{FriedmannC}
    H^2=\frac{\kappa}{3}\left(\rho+V\right).
\end{align}

\subsubsection{Evolution of first order perturbations}
We now consider a perturbed Universe 
whose background is the one of the previous section, and present the first order differential equations for the linearly perturbed CGI variables in~\eqref{list} where, in case of dust, $P^\mu=0$. The shear $\sigma_{\mu\nu}$ and vorticity $\omega_{\mu\nu}$ are not zero, but they decouple from the dynamics because their contribution in~\eqref{eq:Raychaudhuri} and~\eqref{eq:hamiltonian} is quadratic. The vorticity is irrelevant for scalar perturbations. The shear has a scalar contribution which can be obtained from other variables, in particular the curvature perturbation, see \cite{Bruni:1992dg,M.Bruni_etal_2014_Mar}.

Taking the spatial gradient of the continuity equation~\eqref{continuity2}, after some manipulation for which we refer to \ref{app:B}, we obtain the first order differential equation for $\Delta_\mu$:
\begin{align}\label{eq:D1}
    \dot{\Delta}_\mu&=-Z_\mu+\frac{Q}{\rho}\Delta_\mu-\frac{a}{\rho}\left(3H+\frac{Q}{\rho}\right)V_\mu-\frac{a}{\rho}\leftindex^{(3)}\nabla_\mu Q.
\end{align}
With the same procedure,  we obtain the evolution equation for the vacuum density spatial gradient from \eqref{continuity1},
\begin{align}\label{eq:V1}
  \dot{V}_\mu=\frac{Q}{\rho}V_\mu-HV_\mu+\leftindex^{(3)}\nabla_\mu Q.
\end{align}
From this and from \eqref{CGIvacuum1} it is clear that in the case of zero momentum transfer, $V^\mu=f^\mu=0$, the energy exchange 
is homogeneous in the matter rest frame, $\leftindex^{(3)}\nabla_\mu Q=0$. Analogously, starting from the Raychaudhuri equation~\eqref{eq:Raychaudhuri} and taking its spatial gradient (see \ref{app:B} for details) we find:
\begin{align}\label{eq:Z1}
    \dot{Z}_\mu=-2HZ_\mu-\frac{\kappa\rho}{2}\Delta_\mu-\frac{\kappa a}{2}V_\mu+\frac{a}{\rho}\leftindex^{(3)}\nabla^2 V_\mu,
\end{align}
where the spatial Laplacian is defined as $\leftindex^{(3)}\nabla^2\equiv\leftindex^{(3)}\nabla_\mu \leftindex^{(3)}\nabla^\mu$. The Hamiltonian constraint~\eqref{eq:hamiltonian} at first order in perturbations reads:
\begin{align}\label{eq:R0}
    R_\mu=-4a^2HZ_\mu+2\kappa a^2\rho\Delta_\mu+2\kappa a^3 V_\mu,
\end{align}
and, differentiating $R_\mu$ w.r.t. proper time, substituting~\eqref{eq:D1} and~\eqref{eq:Z1} and the below
after some algebra we obtain:
\begin{align}
    \dot{R}_\mu=-\frac{4a^3H}{\rho}\leftindex^{(3)}\nabla^2 V_\mu.\label{eq:R1}
\end{align}
Therefore, when $V^\mu=0$, the curvature perturbation is constant in time. 

\subsection{Geodesic CDM and a pure energy exchange}
From now on, instead of referring to the perfect fluid energy density as $\rho$ we use $\rho_{\rm m}=\rho_{\rm b}+\rho_{\rm c}$ the total matter energy density, namely the sum of the baryonic and CDM components. As for CDM, the baryonic component is modelled by a pressureless fluid ($p_{\rm b}=0$). We therefore study a 3-components Universe two of which, CDM and vacuum, are interacting.

We also wish to consider a scenario where the clustering of CDM is only due to gravity, i.e.\ we need to make sure that the CDM particles do not feel a fifth force due to the interaction.
This means that there must be no acceleration acting on the CDM particles; in other words, CDM must remain geodesic.
By examining \eqref{acceleration}, we can see that for the acceleration $a^\mu_{\,\rm c}=0$, the CGI variables on the right-hand side must vanish.

While $P^\mu=0$ for CDM, to ensure $V^\mu$ vanishes we set $f^\mu=0$ in~\eqref{CGIvacuum1}. This implies that the vacuum itself remains homogeneous in the matter rest frame and any interaction between the vacuum and CDM will be a pure energy exchange, also homogeneous in the matter rest frame, with no momentum transfer. By virtue of \eqref{vacuumvelocity}, this also means that the vacuum 4-velocity equals the CDM one, i.e. $\check{u}^\mu=u^\mu_{\,\rm c}$.

The geodesic CDM assumption implies that the sound speed, $c_s^2$, of vacuum plus matter density perturbations is zero, meaning that the Jeans length vanishes and matter, affected by the vacuum--CDM interaction, clusters on all scales as in standard $\Lambda$CDM \citep{Martinelli:2019dau}.

It is important to stress that this geodesic interaction is an approximation valid only at early times and on large scales. Via the imposition of the geodesic condition, the CDM 4-velocity defines a potential flow, i.e. $u^\mu_{\rm \,c}\propto\nabla^\mu V$, which gives rise to irrotational CDM. At late times, when not only scalar but also vector perturbations are relevant for structure formation, it is expected that non-linear structure growth leads to vorticity~\citep{Pueblas:2008uv,Bruni:2013mua}. Therefore, the assumption of a pure energy exchange, which allows CDM to remain geodesic, must fail below some length scale. If irrotational CDM were present at all scales, distinctive observational features, such as the overproduction of supermassive black holes, would have arisen \citep{Sawicki:2013wja}.

While the spatially flat FLRW interacting background remains unaffected by the geodesic CDM condition, this is not the case for linear perturbations. In the CGI framework, the fact that the momentum transfer $f^\mu=0$ forces the vacuum to remain unperturbed, i.e. $V^\mu=0$. This simplifies considerably the way vacuum--CDM interaction enters into the first order differential equations for the CGI variables $\Delta^\mu$, $Z^\mu$ and $R^\mu$. In particular, we show that some care must be taken when choosing the explicit form of the {\it exact} (and not background) energy exchange $Q$.

\subsubsection{Evolution of first order perturbations}
In the following and for the rest of this paper we will write the comoving fractional total matter density spatial gradient as $\Delta_{\rm m}^\mu$. In addition, we make the assumption that the baryonic component is comoving with CDM at all times. In the CGI language this results in~$\Theta_{\rm b}=\Theta_{\rm c}=\Theta$, meaning that the 4-velocities of baryons and CDM coincide with the one of the perfect fluid, which is now the total matter 4-velocity, i.e. $u^\mu_{\rm m}=u^\mu$.

As a consequence of the geodesic condition $V^\mu=0$, in turn this implies that
\begin{align}\label{eq:SpatQ}
    \leftindex^{(3)}\nabla^\mu Q=HV^\mu-\frac{Q}{\rho_{\rm m}}V^\mu+\dot{V}^\mu=0,
\end{align}
and the dynamical first order differential equations~\eqref{eq:D1},~\eqref{eq:Z1} and~\eqref{eq:R1} reduce to the system of ordinary differential equations (ODEs), cf.~\cite{Kaeonikhom:2022ahf,Salzano:2021zxk}:
\begin{align}
    \dot{\Delta}^\mu_{\rm m}&=-Z^\mu+\frac{Q}{\rho_{\rm m}}\Delta^\mu_{\rm m},\\
    \dot{Z}^\mu&=-2HZ^\mu-\frac{\kappa\rho_{\rm m}}{2}\Delta^\mu_{\rm m},\\
    \dot{R}^\mu&=0.
\end{align}
If we express the linearised Hamiltonian constraint in~\eqref{eq:R0} as a function of $Z^\mu$,
\begin{equation}
    Z^\mu=-\frac{1}{4a^2 H}R^\mu+\frac{\kappa\rho_{\rm m}}{2H}\Delta^\mu_{\rm m},
\end{equation}
we can use the above to write a first order inhomogeneous differential equation for $\Delta_{\rm m}^\mu$,
\begin{align}\label{1ODE}
    \dot{\Delta}^\mu_{\rm m}=\left(\frac{Q}{\rho_{\rm m}}-\frac{\kappa\rho_{\rm m}}{2H}\right)\Delta^\mu_{\rm m}+\frac{1}{4a^2H}R^\mu,
\end{align}
or alternatively, by differentiating with respect to cosmic time, a more standard second order homogeneous differential equation,
\begin{align}\label{2ODE}
    \ddot{\Delta}^\mu_{\rm m}&+\left(2H-\frac{Q}{\rho_{\rm m}}\right)\dot{\Delta}^\mu_{\rm m} \nonumber \\
    &- \left[\frac{\rho_{\rm m}\left(\dot{Q}+5HQ\right)+Q^2}{\rho_{\rm m}^2}+\frac{\kappa}{2}\rho_{\rm m}\right]\Delta^\mu_{\rm m}=0.
\end{align}
The growth of matter perturbations and hence the formation of large-scale structures in the (geodesic CDM) interacting vacuum scenario is enclosed in the growing mode of the above second order ODE. Such a solution is crucially connected with familiar observable quantities in cosmology such as the growth rate $f$ and $\sigma_8$, the amplitude of the matter power spectrum on a scale of $8 \mathrm{h}^{-1}$ Mpc.

\subsubsection{A specific type of energy transfer}\label{sec:ThetaF}
If, in addition to the geodesic condition for which $V_\mu=0$, we assume an interaction of the type $Q\propto\Theta f(V)$, with $f(V)$ a generic function of the vacuum energy density, we find that 
\begin{align}
    \leftindex^{(3)}\nabla_\mu Q\propto\leftindex^{(3)}\nabla_\mu\left[\Theta f(V)\right]=\frac{f(V)}{a}Z_\mu+3Hf^\prime(V)V_\mu,
\end{align}
where the prime denotes a derivative w.r.t. $V$. Then, given the geodesic condition and~\eqref{eq:V1}, the comoving expansion spatial gradient vanishes, $Z_\mu=0$. From~\eqref{eq:Z1},~\eqref{eq:D1} and~\eqref{eq:R0} it follows that we end up with~$\Delta^\mu_{\rm m}=Z_\mu=R_\mu=0$, i.e. with an unperturbed Universe.

However, this does not imply that in the geodesic CDM framework models with a background energy exchange of the form $\bar{Q}\propto H f(\bar{V})$ are unphysical. In fact, such a background form could come from a covariant $Q\propto \Theta f(V)+F(X)$, with $F(X)$ being a function of any scalar $X$ which is built from generic tensorial quantities vanishing in the background, like the CGI variables listed in~\eqref{list}. In this manner, even though $\leftindex^{(3)}\nabla_\mu Q$ and $V_\mu$ are zero, this does  not automatically imply that $Z_\mu=0$, which would lead to an unperturbed Universe. For example, the interacting linear vacuum model that we will later study in section~\ref{sec:method} could be characterised by the energy transfer: 
\begin{align}
    Q=\frac{q}{3}\Theta V+F(X),
\end{align}
which in the background reduces to $Q=qHV$.

We do not intend to speculate further on  possible choices for the function $F(X)$, our intention here being only to underline the physical consistency of this class of geodesic CDM models at first perturbative order. 
In the next sections, after re-expressing the second order ODE~\eqref{2ODE} as a function of the scale factor $a$, we proceed to connect 
the CGI matter perturbation variable
with standard observational probes of structure growth.


\subsubsection{Aggregation of matter}
Following~\cite{Ellis:1990gi,Bruni:1992dg,Dunsby:1991xk}, we now perform a decomposition of $\Delta^\mu_{\rm m}$, applying the spatial gradient to this vector, along the same lines of what we have done for the 4-velocity of a perfect fluid in \eqref{spatial-u-split}; this yields 
\begin{align}\label{eq:trace}
    a\leftindex^{(3)}\nabla^\nu\Delta^\mu_{\text{m}}\equiv\Delta^{\mu\nu}_{\text{m}}=\Omega^{\mu\nu}_{\text{m}}+\Sigma^{\mu\nu}_{\text{m}}+\frac{1}{3}\Delta_{\text{m}} h^{\mu\nu}.
\end{align}
In analogy with the $\leftindex^{(3)}\nabla_\nu u_\mu$ splitting, $\Omega^{\mu\nu}_{\rm m}$ stands for the anti-symmetric part of $\Delta^{\mu\nu}_{\rm m}$, representing a spatial variation of density due to vorticity, $\Sigma^{\mu\nu}_{\rm m}$ for the symmetric traceless part which encodes anisotropy and $\Delta_{\rm m}$ for the trace part, which is directly related to the spherically symmetric spatial variation of density where matter accumulates. This latter quantity is associated with the spatial aggregation of matter which we might expect to reflect the existence of high density structures in the Universe. 

General inhomogeneities will have all three of these components -- for example, there could exist aggregation of matter ($\Delta_{\rm m} > 0$) in a deformed structure ($\Sigma^{\mu\nu}_{\rm m}\neq0$) which has turbulence ($\Omega^{\mu\nu}_{\rm m}\neq0$). In this work we focus on the pure aggregation of matter and therefore from this point on we will neglect the vorticity and anisotropy components.


Using the linear identities detailed in \ref{app:C}, we can rewrite \eqref{2ODE} in terms of the trace of $\Delta^{\mu\nu}_{\rm m}$. In addition, the resulting ODE can be rephrased in terms of the scale factor through $\rm{d}/\rm{d}t=\mathcal{H}\rm{d}/\rm{d}a$ where $\mathcal{H}\equiv aH$ is the conformal Hubble factor; this yields
\begin{align}\label{Da}
    \frac{\rm{d}^2}{\rm{d}a^2}\Delta_{\rm m}&+a^{-1}\left(3+g-\frac{3}{2}\Omega_{\rm m}\right)\frac{\rm{d}}{\rm{d}a}\Delta_{\rm m} \nonumber \\
    &+ \left[ \left(a\mathcal{H}\right)^{-1}\frac{\rm{d}}{\rm{d}a}\left(g\mathcal{H}\right)+a^{-2}\left(g-\frac{3}{2}\Omega_{\rm m}\right)\right]\Delta_{\rm m}=0,
\end{align}
where $\Omega_{\rm m}\equiv\frac{\kappa\rho_{\rm m}}{3H^2}$ and $g\equiv-\frac{Q}{H\rho_{\rm m}}$ are the dimensionless total matter density and  dimensionless interaction parameters respectively.

It is standard practice to write the general solution of~\eqref{Da} as a linear combination of growing and decaying modes ($D_{\mathrm{m}}^{+}$ and $D_{\mathrm{m}}^{-}$). The decaying mode is the solution of the homogeneous part of~\eqref{1ODE} whereas the growing mode is the solution driven by the constant curvature perturbation source term $R_\mu$. We disregard the contribution of the decaying mode since it is negligible at late times.

\subsubsection{Effect of perturbations on  observable quantities}
The above discussion takes on a more familiar form with the imposition of the comoving-orthogonal gauge which, for a geodesic CDM interaction, automatically implies a synchronous frame (we show this explicitly in \ref{app:D}). This leads to the CGI quantity $\Delta_{\rm m}$ reducing to the matter density contrast $\delta_{\rm m}$ in that gauge. We can use this result to connect $\delta_{\rm m}$ with an observational probe of structure growth, namely $f \sigma_8(a)$. This quantity describes the rate of change of the amplitude of clustering,
\begin{equation}
    f \sigma_8(a) = \frac{\mathrm{d} \sigma_8}{\mathrm{d} \ln a},
\end{equation}
where the growth rate of cosmic structures $f$ is defined as usual in terms of the linear growing mode $D^{+}_{\rm m}$:
\begin{equation}
    f(a) = \frac{\mathrm{d}\ln D^{+}_{\rm m}}{\mathrm{d} \ln a}. \label{eq:dlnD}
\end{equation}
The parameter
$\sigma_8$ is the root mean square amplitude of matter density fluctuations on a comoving scale of $R = 8 \mathrm{h}^{-1}$ Mpc,
\begin{equation}
    \sigma_{R}^2(a) =  \int_0^\infty \frac{k^3 P_{\rm linear}(k, a)}{2 \pi^2} \, W^2(kR) \, \mathrm{d} \ln k,
\end{equation}
where $P_{\rm linear}(k, a)$ is the linear matter power spectrum and $W(kR)$ is the window function. 
The scale factor dependence of the $\sigma_8$ parameter
can be expressed as 
\begin{align}\label{sigma8a}
    \sigma_8(a)=\sigma_8(1)\frac{D^{+}_{\rm m}(a)}{D^{+}_{\rm m}(1)},
\end{align}
which is a common choice to describe the normalisation of matter density fluctuations.
Due to the contribution of the interaction to the evolution of the matter density contrast \eqref{Da}, the original growth rate which we can call $f_{\rm bare}$ (as defined in \eqref{eq:dlnD}) picks up an additional term \citep{Martinelli:2019dau}, 
\begin{equation}\label{eq:fi}
    f_{\rm i} = f_{\rm bare} + g,
\end{equation}
where $g$ was defined above. This means that observational probes such as redshift space distortions (RSDs; the distortions of galaxy clustering when observed in redshift space due to their peculiar velocities) measure $f_{\rm i}\sigma_8$ rather than $f_{\rm bare} \sigma_8$ in interacting vacuum cosmologies. For notational simplicity, we will refer to $f_{\rm i}$ as $f$ from now on. This is relevant due to our later use of constraints obtained on interacting models using RSD data.

\section{Model choice and analysis method}\label{sec:method}
Up to this point, we have derived the dynamical equations for both background and linear perturbations and introduced some relevant quantities for the study of the growth of cosmic structures in the interacting vacuum scenario, while remaining agnostic about the exact form that the interaction takes. 

We now describe the specific models for the interaction that we choose to study in this paper. With the specific interaction models chosen, we can solve the equations for the background and first order perturbations in each model, and make plots of observable quantities such as $h$ and $f \sigma_8$. With the cosmological behaviour of these models made clear in these plots, we will show situations in which cosmological tensions can or cannot be resolved in the interacting vacuum picture. We remind the reader that from now on we set $\kappa=1$.

\subsection{Model choice}

In a geodesic CDM framework, we consider three interacting vacuum models: the first with a background interaction linear in the vacuum energy density \citep{Wands:2012vg, Salvatelli:2014zta, Martinelli:2019dau, Kaeonikhom:2022ahf}, the second with a background interaction mimicking a generalised Chaplygin gas \citep{Wang:2013qy, Wang:2014xca} and the third with a background interaction inspired by a Shan--Chen fluid \citep{Hogg:2021yiz}. While the generalised Chaplygin gas and Shan--Chen dark energy models were originally proposed as fluid dark energy models \citep{Bento:2002ps, Bini:2013ods}, for the purposes of the comparisons in this work, we express them as interacting vacuum models.

\subsubsection{Interacting linear vacuum model} \label{subsec:LV}
In the interacting linear vacuum (LV) model \citep{Wands:2012vg, Salvatelli:2014zta,Martinelli:2019dau,Kaeonikhom:2022ahf}, the background coupling function is given by
\begin{equation}
    Q = qHV \label{LV},
\end{equation}
where the magnitude of the constant dimensionless coupling parameter $q$ controls the strength of the vacuum--CDM interaction and its sign dictates the direction of the energy transfer.
Since both the vacuum energy density, $V$, and the Hubble function, $H$, are positive quantities, we can see that for $q>0$ the vacuum energy density grows and, due to total energy conservation, CDM decays; in other words, we have an energy transfer from CDM to vacuum. Conversely, for $q<0$ the vacuum energy density decays and CDM grows; the non-interacting or \LCDM limit occurs for $q=0$.

Integrating the continuity equations \eqref{continuity1} and \eqref{continuity2} with today initial conditions, $a=1$, we obtain the following analytical solutions for the CDM and vacuum energy densities:
\begin{align}
    \rho_\text{c}(a)&=3H^2_0\left[\Omega_{\rm c, 0}a^{-3}+\Omega_{\rm V, 0}\frac{q}{3+q}\left(a^{-3}-a^{q}\right)\right],\label{rhoq}\\
    V(a)&=3H^2_0\Omega_{\rm V, 0}a^q,\label{Vq}
\end{align}
where $H_0$ is the Hubble parameter and $\Omega_{\rm c, 0}$ and $\Omega_{\rm V, 0}$ are the current (i.e. $z=0$) values of the  dimensionless density parameters for CDM and vacuum. 

From the Friedmann constraint equation \eqref{FriedmannC}, we derive analytical expressions for the reduced dimensionless Hubble function $h(a)$, the dimensionless total matter density $\Omega_{\rm m}(a)$ and the dimensionless interaction function $g(a)$:
\begin{align}
h(a)&=h_0\left[\frac{3\left(1-\Omega_{\rm m, 0}\right)a^{3+q}+3\Omega_{\rm m, 0}+q}{\left(3+q\right)a^3}\right]^{1/2},\\
\Omega_{\rm m}(a)&=\frac{3\Omega_{\rm m, 0}+q-q(1-\Omega_{\rm m, 0})a^{3+q}}{3\Omega_{\rm m, 0}+q+3(1-\Omega_{\rm m, 0})a^{3+q}}\label{qOmega},\\
g(a)&=-\frac{\left(1-\Omega_{\rm m}(a)\right)}{\Omega_{\rm m}(a)}q,\label{eq:gq}
\end{align}
where $\Omega_{\rm m, 0}=\Omega_{\rm b, 0}+\Omega_{\rm c, 0}$ and $h_0=\frac{H_0}{\text{100 km s}^{-1}\text{Mpc}^{-1}}$. An Einstein--de Sitter cosmology is recovered at early times, i.e. for $a\ll1$, when $\Omega_{\rm m}\simeq 1$ and $g\simeq 0$ \citep{Borges:2017jvi}.\footnote{Notice from \eqref{qOmega} that for $q>0$, $\Omega_{\rm m}(a)$ asymptotes to a negative constant value at late times ($a\gg1$). This unphysical limit is a feature of the interacting linear vacuum model that has already been pointed out in previous works \citep{Borges:2017jvi,Martinelli:2019dau}.}

Moving now to first order perturbations, we numerically integrate \eqref{Da} and obtain a solution for the linear matter growing mode $D_{\rm m}^{+}(a)$. The initial conditions are fixed during matter domination, at $a=\frac{1}{1+1000}$, when we have: $\Omega_{\rm m}\simeq1$, $g\simeq0$ and $D_{\rm m}^{+}\simeq a$.

\subsubsection{Interacting generalised Chaplygin gas model} \label{subsec:gCg}
The second model we consider is derived from the generalised Chaplygin gas (gCg) \citep{Kamenshchik:2001cp,Bento:2002ps}, a unified dark matter fluid characterised by the equation of state
\begin{align}\label{equation of stategCg}
    p_{_{\rm gCg}}=-A\rho_{_{\rm gCg}}^{-\alpha}.
\end{align}
Following  \cite{Bento:2004uh}, we think of the interacting CDM and vacuum so that their total energy density, $\rho_{\rm tot}$, is the same as that of the gCg:
\begin{align}
  \rho_{\rm tot} &\equiv  \rho_{_{\rm gCg}}=\rho_{\rm c}+V,\notag\\
    V&=-p_{_{\rm gCg}}.\label{DgCg}
\end{align}
Using \eqref{equation of stategCg} and \eqref{DgCg} we find  
$A=V(\rho_{\rm c}+V)^\alpha$ and, inserting this expression into \eqref{continuity2}, we obtain the background energy transfer function for the interacting generalised Chaplygin gas model \citep{Bento:2004uh,Wang:2013qy,Wang:2014xca}:
\begin{align}\label{eq:Qalpha}
    Q&=3\alpha H\frac{V\rho_{\rm c}}{V+\rho_{\rm c}} 
\end{align}
The sign and magnitude of the constant dimensionless coupling parameter $\alpha$ denotes the direction and strength of the CDM--vacuum interaction: if  $\alpha>0$ the energy is transferred from CDM to vacuum, vice versa for $\alpha<0$.

From \eqref{equation of stategCg} and \eqref{DgCg} and integrating the continuity equation for the single component generalised Chaplygin gas with initial conditions set at $a=1$, we obtain the following analytical solution for the total energy density:
\begin{align}\label{eq:rhogCg}
    \rho_{\rm tot}(a)=& \; 3H_0^2\left(\Omega_{\rm c, 0}+\Omega_{\rm V, 0}\right)^{\frac{\alpha}{1+\alpha}} \nonumber \\
    &\times \left[\Omega_{\rm V, 0}+\Omega_{\rm c, 0}a^{-3(1+\alpha)}\right]^{\frac{1}{1+\alpha}}.
\end{align}
From the Friedmann constraint equation \eqref{FriedmannC}, we get
\begin{align}
    h(a)&=h_0\left[\Omega_{\rm b,0}a^{-3}+\left(\Omega_{\rm c,0}+\Omega_{\rm V,0}\right)^{\frac{\alpha}{1+\alpha}} \right. \nonumber \\
    &\times \left. \left(\Omega_{\rm V,0}+\Omega_{\rm c,0}a^{-3(1+\alpha)}\right)^{\frac{1}{1+\alpha}}\right]^{\frac{1}{2}},
\end{align}
and, substituting \eqref{eq:Qalpha} in the dimensionless interaction parameter $g$, we find
\begin{align}\label{ggCg}
    g(a)=-3\alpha\frac{\Omega_{\rm c}(a)}{\Omega_{\rm m}(a)}\frac{\Omega_{\rm V}(a)}{\Omega_{\rm c}(a)+\Omega_{\rm V}(a)}.
\end{align}
By looking at $\Omega_{\rm V}(a)$ explicitly,
\begin{align}
\Omega_{\rm V}(a)&=a^{3(1+\alpha)}\Omega_{\rm V, 0}\left(\Omega_{\rm c, 0}+\Omega_{\rm V, 0}\right)^{\frac{\alpha}{1+\alpha}} \nonumber \\
&\times \left(\Omega_{\rm c, 0}+a^{3(1+\alpha)}\Omega_{\rm V, 0}\right)^{-\frac{\alpha}{1+\alpha}} \nonumber \\
&\times \left[1-\left(\Omega_{\rm c, 0}+\Omega_{\rm V, 0}\right)+\left(\Omega_{\rm c, 0}+\Omega_{\rm V, 0}\right)^{\frac{\alpha}{1+\alpha}} \right. \nonumber \\
&\times \left. \left(\Omega_{\rm c, 0}+a^{3(1+\alpha)}\Omega_{\rm V, 0}\right)^{\frac{1}{1+\alpha}}\right]^{-1},
\end{align}
it is easy to see that a standard matter dominated era is again recovered for $a\ll1$, when $\Omega_{\rm V} \simeq 0$ and $g \simeq 0$. We deal with linear matter density perturbations in the same way as in the interacting linear model.

\subsubsection{Interacting Shan--Chen model} \label{subsec:SC}
The third interacting vacuum model we analyse originates from the non-linear Shan--Chen (SC) equation of state which was originally used to describe matter in lattice kinetic theory~\citep{Shan1993}. This equation of state was then repurposed to describe a dark energy fluid \citep{Bini:2013ods,Bini:2016wqr} and subsequently an interacting vacuum dark energy \citep{Hogg:2021yiz}. The non-linear Shan--Chen equation of state is
\begin{align}
p_{_{\rm SC}}&=\beta\rho_*\left[\frac{\rho_{_{\rm SC}}}{\rho_*}+\frac{G}{2}\psi^2\right], \label{SCeos} \\
\intertext{with}
\psi&=1-\mathrm{e}^{-\alpha\frac{\rho_{\scaleto{\rm SC}{2.5pt}}}{\rho_*}}. \label{SCpsi}
\end{align}
The parameters $\beta$, $G$ and $\alpha$ are all dimensionless and control various aspects of the behaviour of the fluid (see \cite{Hogg:2021yiz} for an overview of the behaviour of the interacting vacuum Shan--Chen model). The parameter $\rho_*$ is a characteristic energy density scale which we keep fixed to the value of the critical density today, $\rho_*=3H_0^2$. This is to ensure that the interaction has maximum effect at late times, but such a model could be adapted to create an early dark energy model by setting this energy scale to, for example, the critical density at recombination.
We build the corresponding interacting vacuum model by treating CDM as pressureless dust and vacuum energy as a SC dark energy, i.e. $V=\rho_{_{\rm SC}}$. From the SC continuity equation
\begin{align}\label{eq:continuitySC}
    \dot{\rho}_{_{\rm SC}}=-3H\left(\rho_{_{\rm SC}}+p_{_{\rm SC}}\right),
\end{align}
substituting \eqref{SCeos} and \eqref{SCpsi}, we obtain the background energy transfer parameterisation that enters in \eqref{continuity1} and \eqref{continuity2},
\begin{equation}\label{SC}
Q=3qH\left[\left(1+\beta\right)V+\frac{\beta G}{2}\rho_*\left(1-\mathrm{e}^{-\alpha\frac{V}{\rho_*}}\right)^2\right],
\end{equation}
where, as in the previous models we have described, the dimensionless coupling strength parameter, $q$, controls the overall strength and direction of the interaction.\footnote{Note that in \eqref{SC} we have absorbed a minus sign in the $q$ parameter. In this manner, when $\beta=0$, the SC model reduces to an interacting linear vacuum.}

Due to the complexity of this parameterisation, it is not possible to obtain analytical solutions for this model. Instead, we numerically solve the dynamical equations for both the background and first order perturbations, setting the initial conditions as in the other two discussed models.

\subsection{Analysis method} \label{sec:analysis}
Having solved the equations for the background and first order perturbations in each of the models described above, we plot various observables to understand the different cosmologies produced by each interacting model. To do this, we need to choose some reference values for the parameters, such as $q$, so that a wide range of possible cosmological effects can be seen in each model. We then use results from the literature to select values of the model parameters which are favoured by observable data, and plot the same cosmological quantities again, in order to make a qualitative comparison. The theoretical behaviour of the models allows us to identify which parts of parameter space may or may not lead to resolutions of the cosmological tensions in each model, whereas the plots using parameter values informed by data will show us how these models actually fare when confronted by observations.

\subsubsection{Initial conditions}
For what we will refer to as the theory plots, the initial conditions of \eqref{continuity1}, \eqref{continuity2} and \eqref{Da} are fixed during matter domination, at $a=\frac{1}{1+1000}$, when the vacuum--CDM interaction is negligible for the models we consider. We do this by evolving a reference \LCDM cosmology backwards in time and fixing the initial conditions at the aforementioned redshift. This amounts to turning the interaction on starting from a common \LCDM cosmology in the past, characterised by the parameters 
$h_0=0.70$, $\Omega_{\rm b, 0}=0.05$, $\Omega_{\rm c, 0}=0.25$ and $\sigma_8=0.80$. The interaction will then affect the future evolution in each model depending on the sign and magnitude of the corresponding coupling strengths.

For what we will refer to as the data plots, the initial conditions are fixed to the current i.e. $a=1$ values as constrained by observational data. We specifically use the values from \cite{Kaeonikhom:2022ahf} for the linear vacuum, \cite{Wang:2014xca} for the generalised Chaplygin gas, and \cite{Hogg:2021yiz} for the Shan--Chen model. We list the numerical values for the parameters in tables in each subsection.

\section{Results}\label{sec:results}
In this section, we present our results for the three interacting vacuum models we consider. We show the dynamical evolution of cosmological quantities of interest: $h$, $\Omega_{\rm m}$, $\sigma_8$ and $f\sigma_8$. For each model, we present both the plots made by setting the model parameters to some arbitrary reference values (theory plots) and the plots made by setting the parameters to values obtained from observational data constraints (data plots). We analyse the effects that the vacuum--CDM interaction has on the late-time dynamics of the relevant cosmological functions by examining their deviations from the non-interacting \LCDM limit.

\subsection{Interacting linear vacuum} 
\label{subsec:LV_plots}
We begin with the interacting linear vacuum, $Q=qHV$.

\subsubsection{Theory plots}\label{sec:LVtheory}

\begin{figure*} 
\centering 
\includegraphics[width=0.49\textwidth]{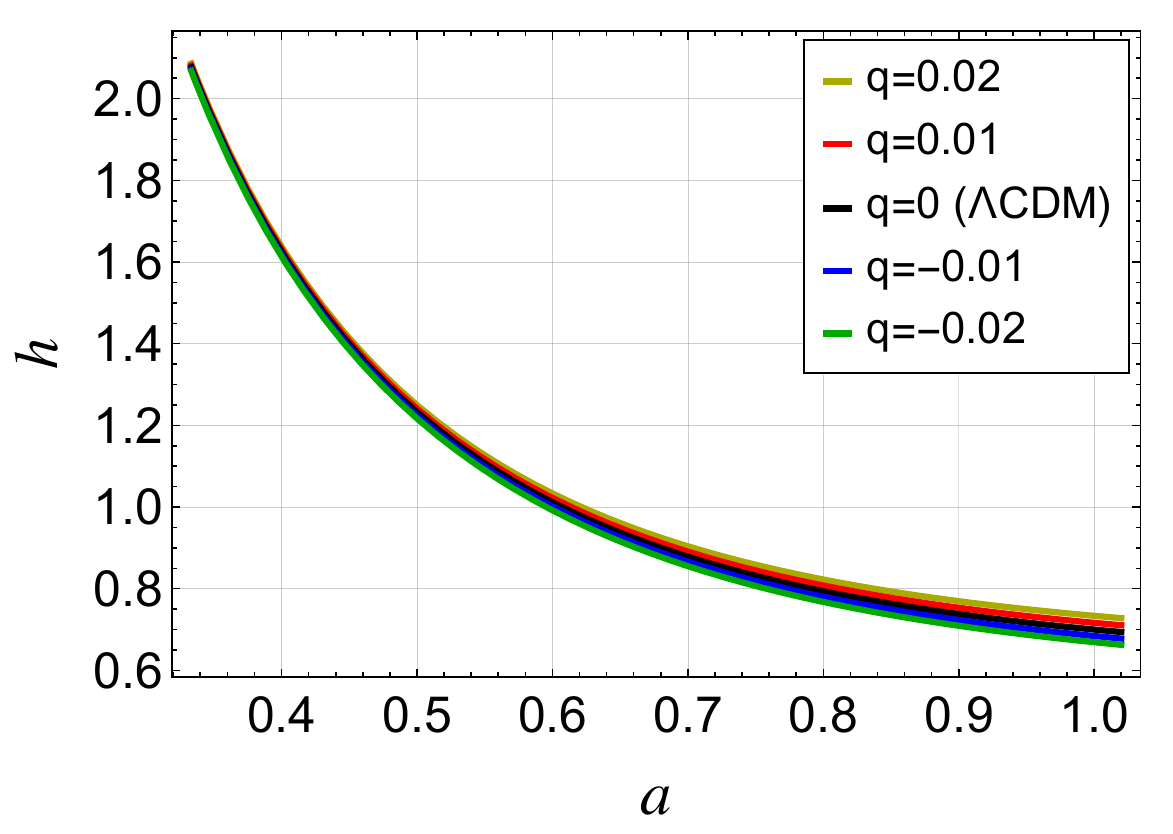}
\includegraphics[width=0.49\textwidth]{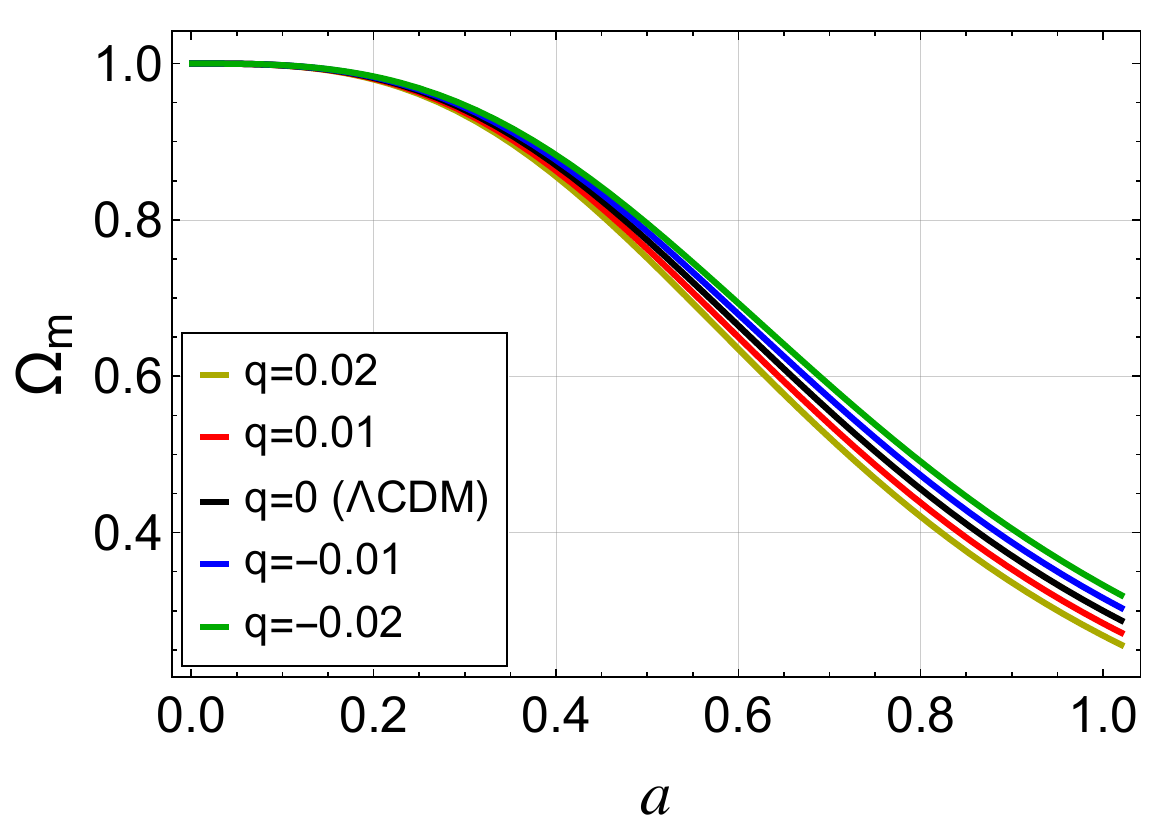}\\
\includegraphics[width=0.49\textwidth]{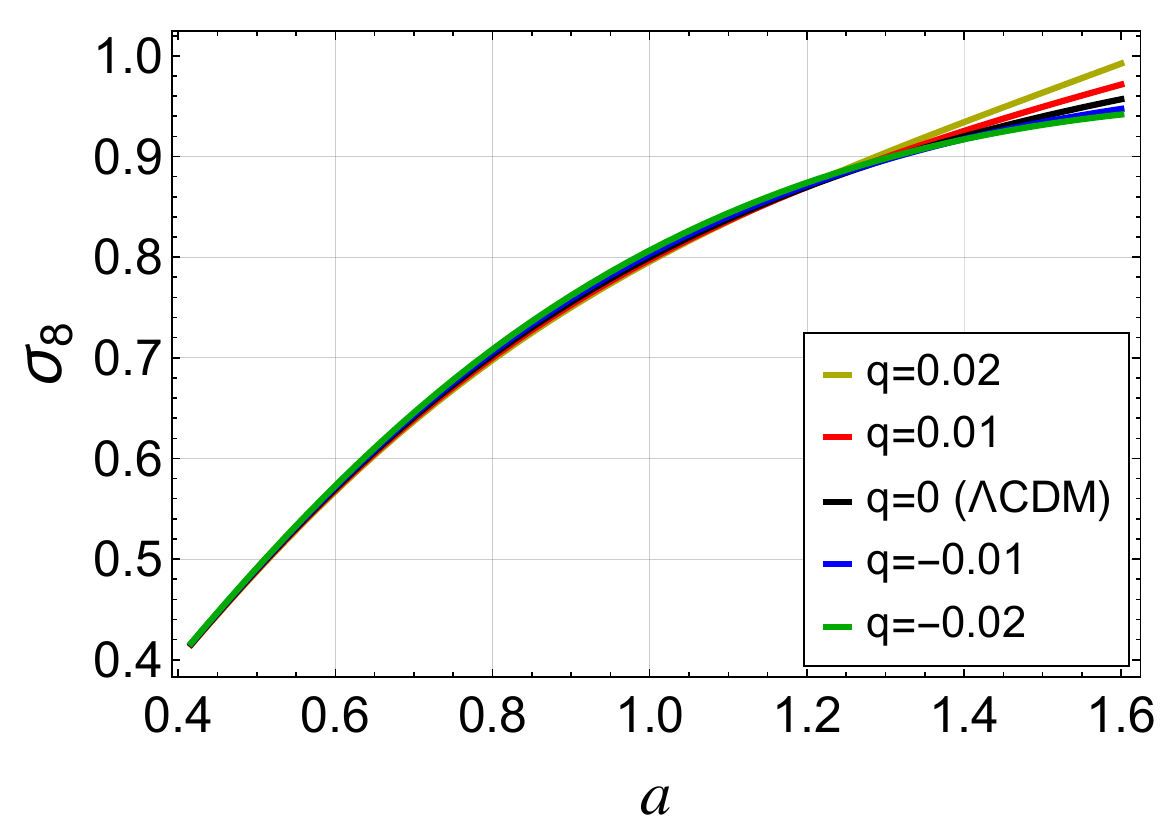}
\includegraphics[width=0.49\textwidth]{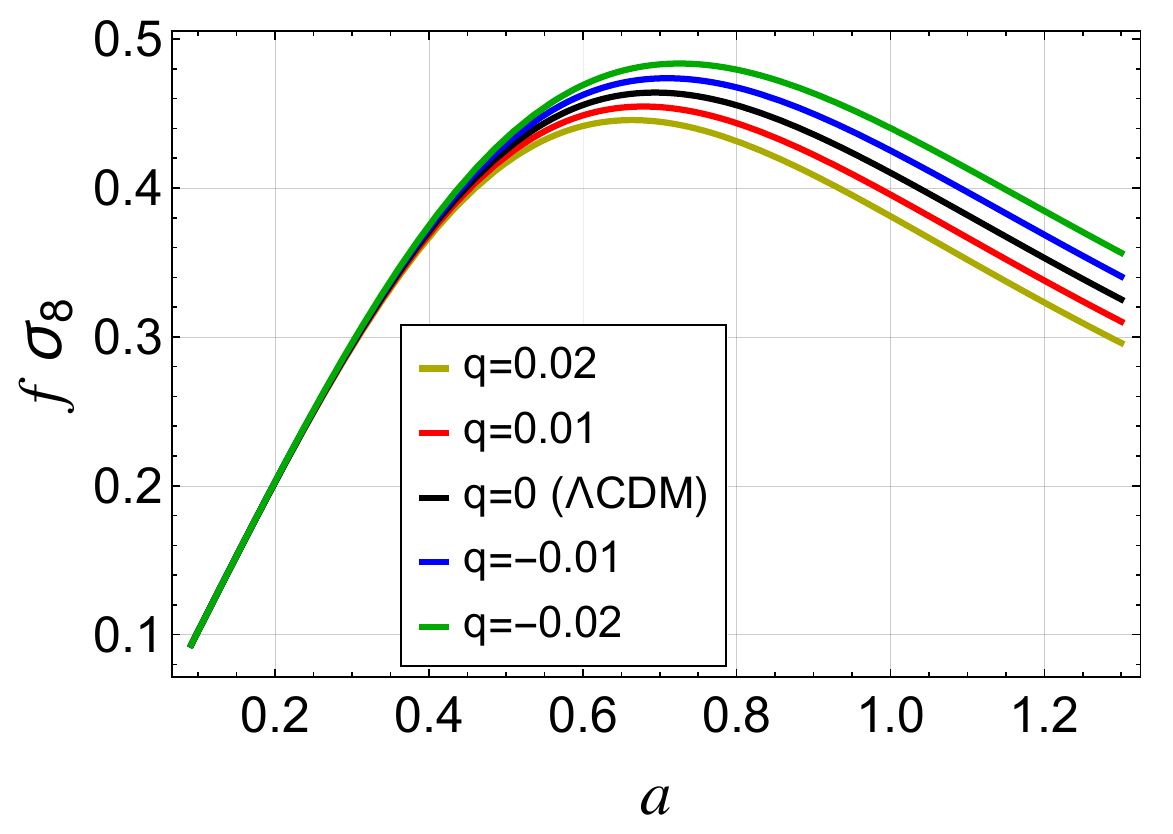}
\caption{Clockwise from top left: $h(a)$, $\Omega_{\rm}(a)$, $f \sigma_8(a)$ and $\sigma_8(a)$ in the interacting linear vacuum model. Five different values of the coupling strength $q$ are shown: $q=0.02$ (yellow), $q=0.01$ (red), $q=0$ (black; i.e. the \LCDM limit), $q=-0.01$ (blue), $q=-0.02$ (green).}
\label{fig:LV_theory}
\end{figure*}

In figure \ref{fig:LV_theory} we show the evolution of $h(a)$, $\Omega_{\rm m}(a)$, $\sigma_8(a)$ and $f\sigma_{8}(a)$, using example values of the coupling strength $q$ between $-0.02$ and $0.02$. The \LCDM cosmology (black), characterised by $h_0=0.70$, $\Omega_{\rm b, 0}=0.05$, $\Omega_{\rm c, 0}=0.25$ and $\sigma_8=0.80$, is given by $q=0$.

From the top two panels of this figure we can see that for a growing vacuum, i.e. for $q>0$, we obtain a smaller matter density, $\Omega_{\rm m}(a)$, and a faster expansion rate, $h(a)$, compared to \LCDM. This is in line with our expectations for cosmologies where vacuum grows at the expense of CDM. We can see that the opposite behaviour occurs for a decaying vacuum, i.e. for $q<0$.

From the bottom two panels of this figure we can see that for a growing vacuum $\sigma_8(a)$ is initially suppressed compared to the \LCDM limit, and is then enhanced in the future ($a>1$). In contrast, $f \sigma_8(a)$ is consistently suppressed compared to \LCDM throughout its evolution. The opposite is true (in both cases) for a decaying vacuum.

At first glance, we may have expected to find that in a decaying vacuum cosmology the clustering of matter would be enhanced due to the growth of CDM at the expense of the vacuum, but instead we find the opposite. This is because the growth of structures is regulated by the friction term in the differential equation for the evolution of the matter density contrast, \eqref{Da}, which enters into the expression for $\sigma_8(a)$ by means of \eqref{sigma8a}. The smaller the friction term, the faster the matter perturbations will grow. Taking the case of the decaying vacuum, when $a<1$, the friction is slightly smaller than in \LCDM, causing perturbations to grow faster while, at late times, this term is increased compared to \LCDM, hence perturbations are suppressed.
The opposite is true for the case of a growing vacuum, and helps us to understand the counter-intuitive behaviour of $\sigma_8(a)$.

The evolution of $f\sigma_{8}(a)$ can be understood by looking at a first order approximation for the growth rate $f\approx\Omega_{\rm m}^{\gamma}$, where the parameter $\gamma>0$ for the range of $q$ under consideration \citep{Borges:2017jvi}. The fact that $f$ goes like the dimensionless matter density and that $\sigma_8\sim0.80$ for $q\sim10^{-2}$ and around $a\sim1$, implies that the $f\sigma_8$ trend will not be very different from the trend in $\Omega_{\rm m}$ at late times. In particular, the evolution of $f \sigma_8(a)$ for a given value of $q$ will not cross its evolution for another value, unlike in the case of $\sigma_8(a)$.

\begin{figure*} 
\centering
\includegraphics[width=0.49\textwidth]{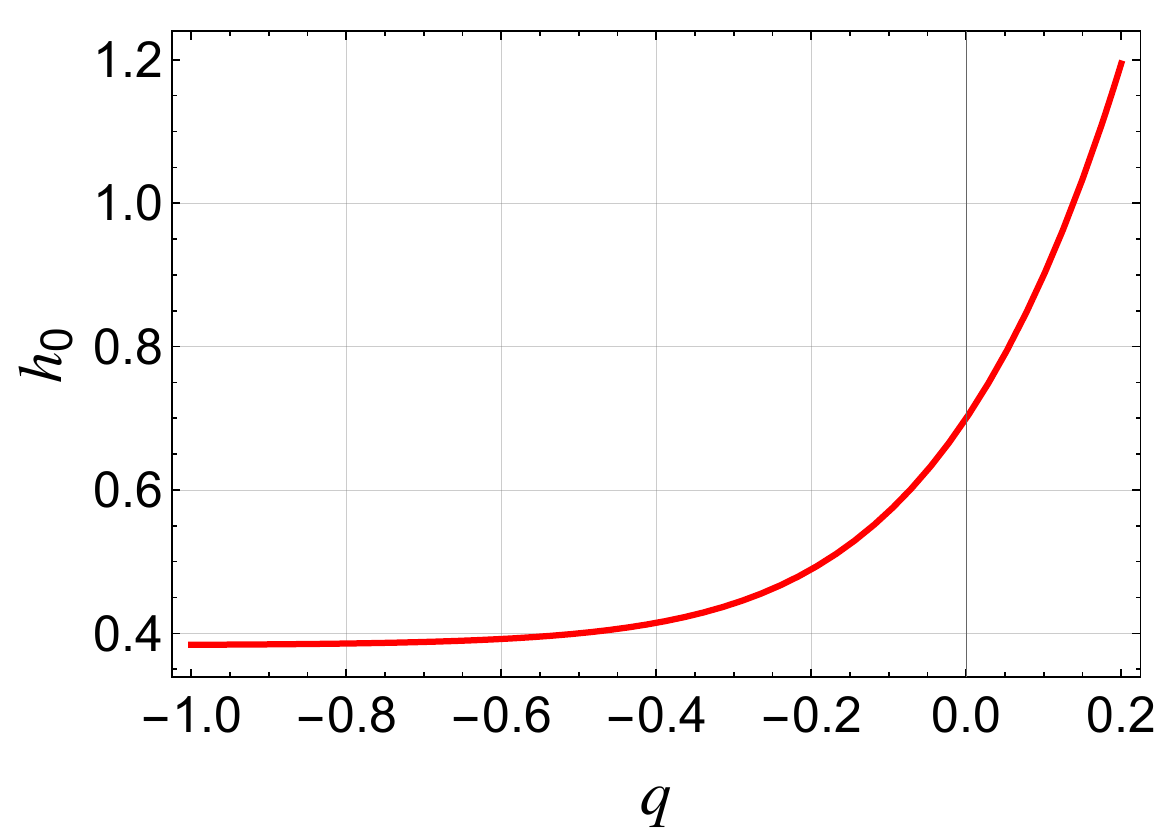}
\includegraphics[width=0.49\textwidth]{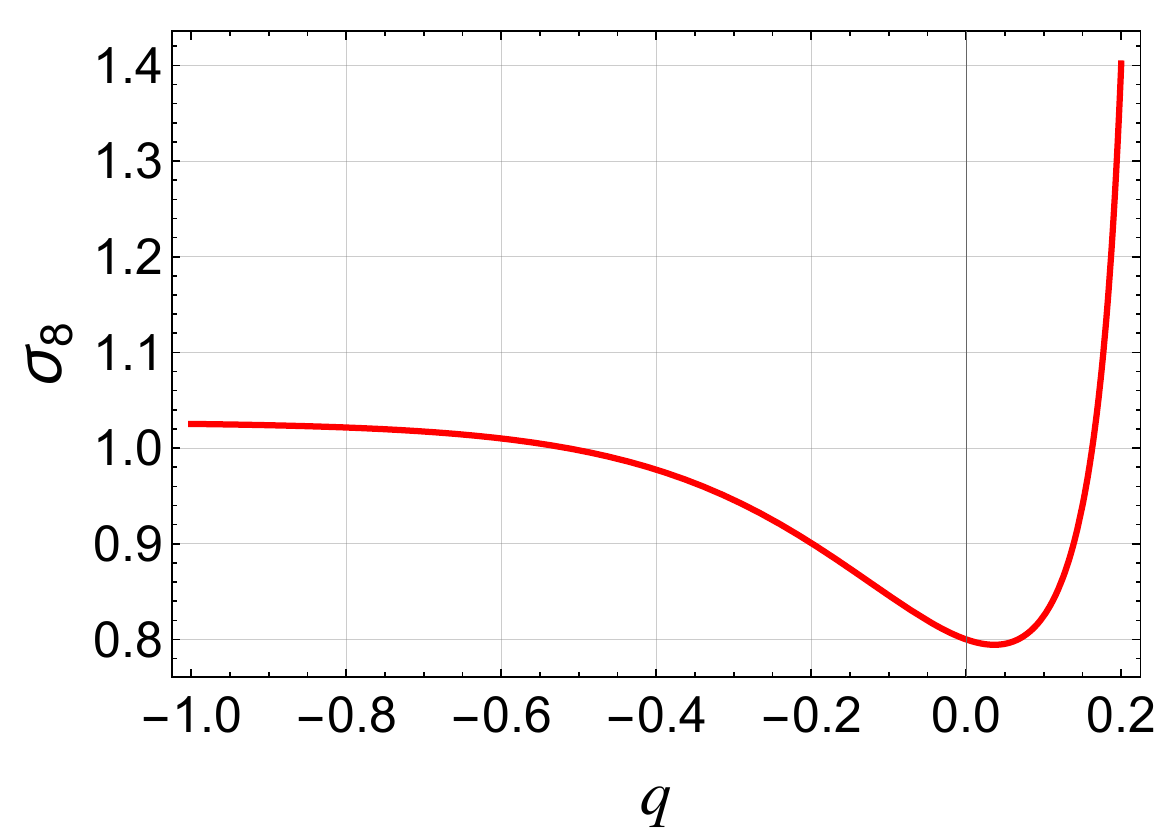}
\caption{\textbf{Left:} how $h_0$ changes as a function of $q$. \textbf{Right:} how $\sigma_8\equiv\sigma_8(z=0)$ changes as a function of $q$. The vertical line in each panel at $q=0$ indicates the \LCDM limit.}
\label{fig:LV_today}
\end{figure*}

We can visualise the results in another way. In figure \ref{fig:LV_today}, we show how the values of $h_0\equiv h(z=0)$ and $\sigma_8\equiv \sigma_8(z=0)$ change with the coupling strength $q$. If we take for example $q=0.1$, we can see from the left-hand panel of the figure that $h_0$ for this value of $q$ is larger than $h_0(q=0)$, namely the \LCDM reference value, hence allowing the $H_0$ tension to be resolved. However, the right-hand panel shows us that $\sigma_8(q)$ has a minimum around the \LCDM limit, therefore, both negative and positive values of $q$ will enhance the LV $\sigma_8$ with respect to the \LCDM one. The only exception to this is between $q=0$ and $q\approx0.067$, for which a slight decrease of $\sigma_8$ compared to $\Lambda$CDM is theoretically allowed. Regardless, we consider its minimum value, $\sigma_8(q\approx0.036)\approx0.794$, too close to \LCDM to concretely signal a parameter space region corresponding to an observable reduction in the $\sigma_8$ tension. We conclude that, with a coupling strength that alleviates the $H_0$ tension, the $\sigma_8$ tension will be unavoidably exacerbated in the interacting linear vacuum model.

\subsubsection{Data plots}\label{sec:LVdata}
The above discussion used arbitrary values of the coupling strength to elucidate the cosmological behaviour in the interacting linear vacuum model. We now present the evolution of the same cosmological parameters when the initial conditions are fixed using observational data. We use the results reported in \cite{Kaeonikhom:2022ahf}, which were obtained with a combination of the Planck 2018 measurements of the CMB TTTEEE power spectra \citep{Aghanim:2018eyx}, BOSS and eBOSS, BAO and redshift space distortion data \citep{Alam:2016hwk, Bautista:2020ahg, Gil-Marin:2020bct, Hou:2020rse, Neveux:2020voa}, and the Pantheon SNIa catalogue \citep{Scolnic:2017caz}. We report the specific values we use in table \ref{tab:LV_table}. For each of the panels, the parameters not displayed are kept fixed to their best-fit values listed in the table.

\begin{figure*} 
\centering 
\includegraphics[width=.49\textwidth]{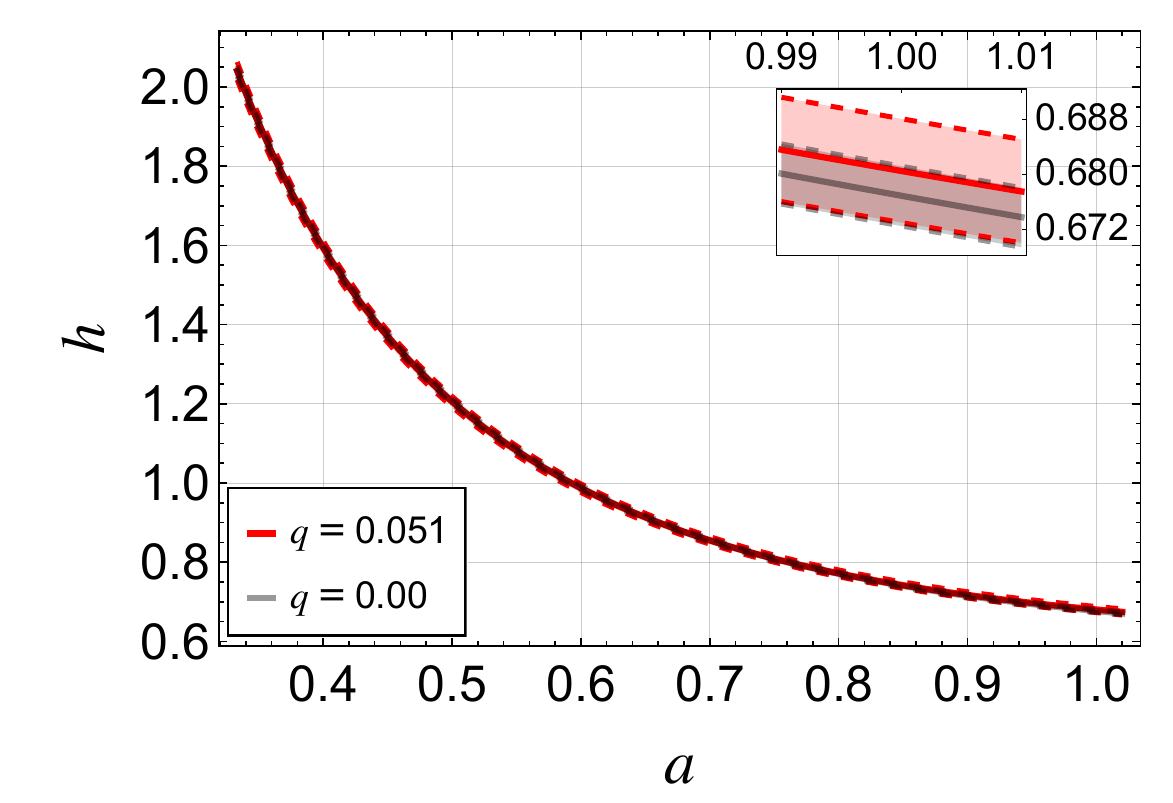}
\includegraphics[width=.49\textwidth]{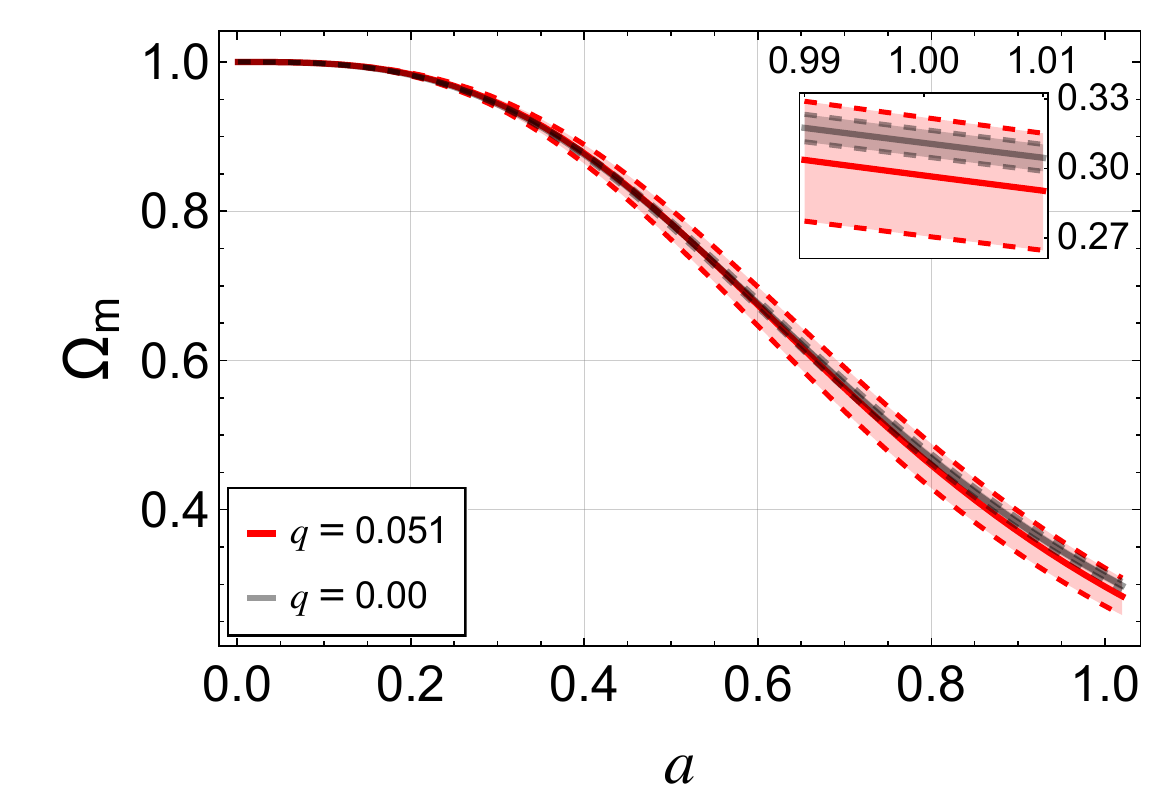}
\includegraphics[width=.49\textwidth]{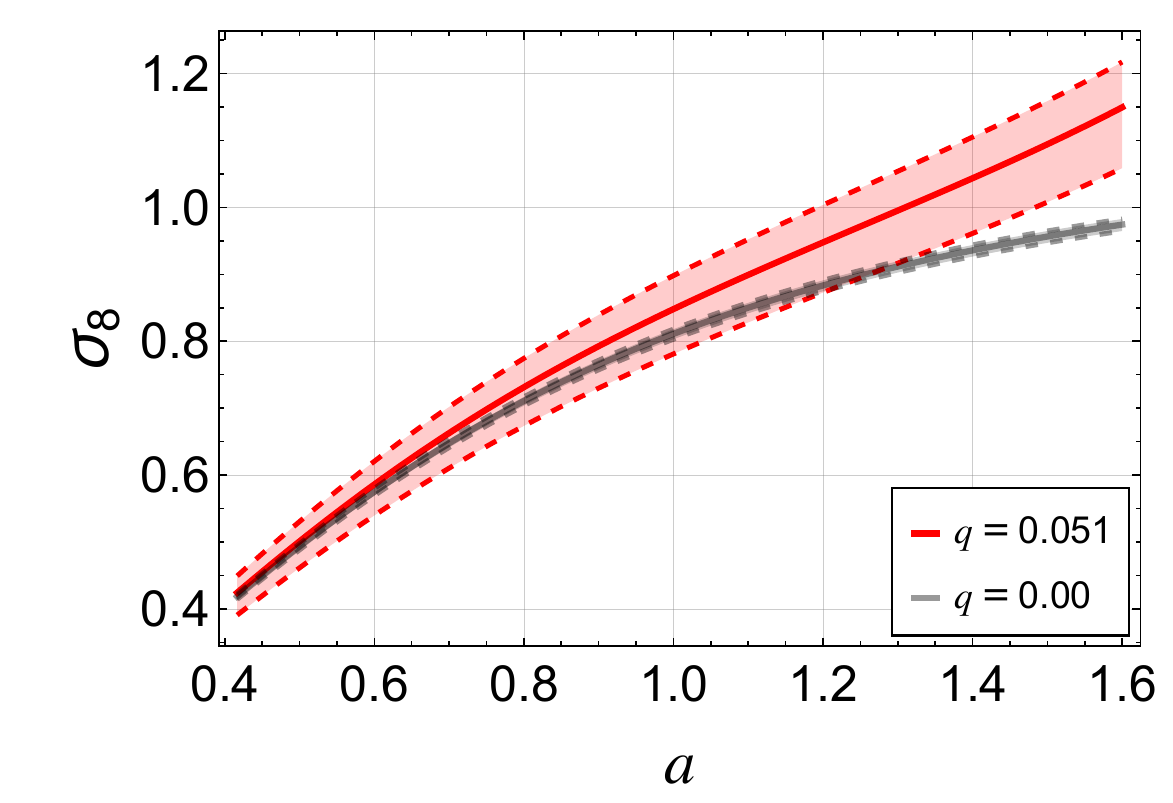}
\includegraphics[width=.49\textwidth]{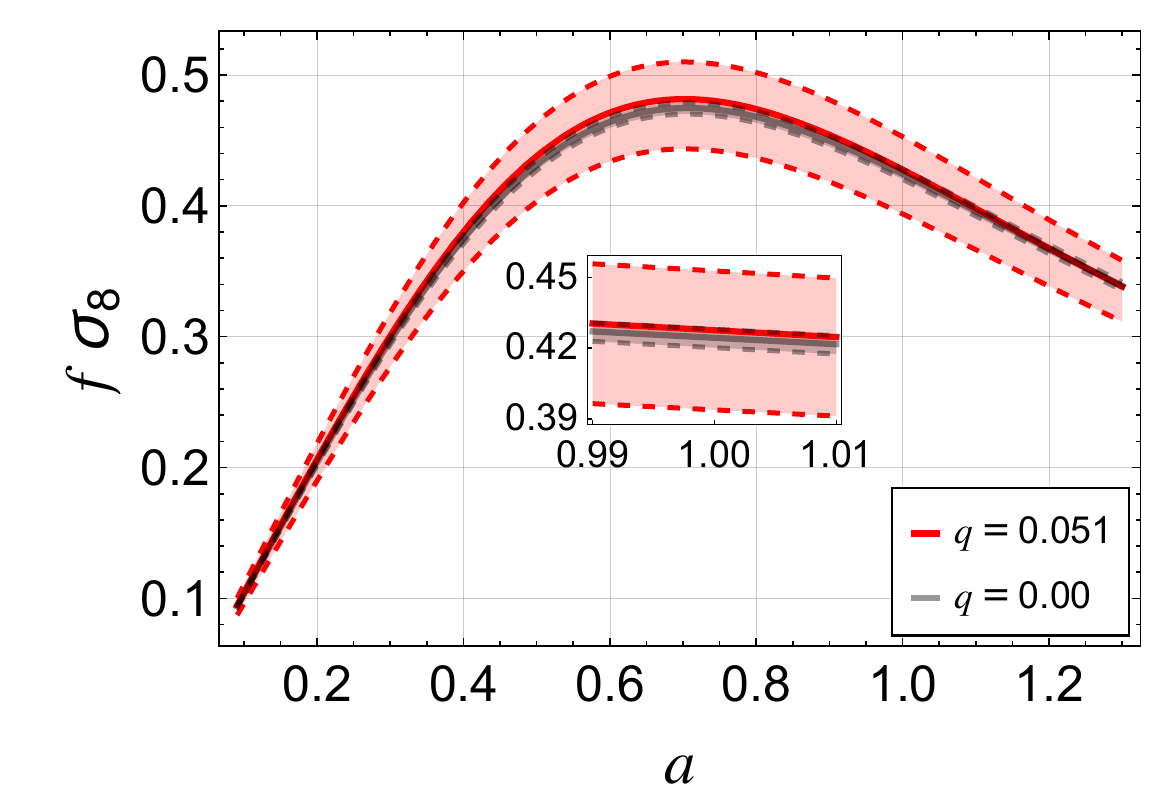}
\caption{Clockwise from top left: $h(a)$, $\Omega_{\rm}(a)$, $f \sigma_8(a)$ and $\sigma_8(a)$ in the interacting linear vacuum model as constrained by data. The grey lines show the \LCDM results and the red lines show the interacting linear vacuum results as obtained in \cite{Kaeonikhom:2022ahf}. The shaded regions represent the $1\sigma$ error bars.}
\label{fig:LV_data}
\end{figure*}

The results of this analysis are shown in figure \ref{fig:LV_data}, where we have extrapolated the $1 \sigma$ error bars given on the parameters at $z=0$ to all values of $a$ (the same will later apply to the interacting generalised Chaplygin gas and Shan--Chen models). Given that the datasets used to obtain these errors are not continuous throughout the same time period, this means that these uncertainties are likely underestimated outside the ranges of the data. However, they provide a useful guide to the eye.

\begin{table*} 
\centering 
\begin{tabular}{Slcccc} 
\hline
\hline
Model & $q$ & $H_0$ & $\Omega_{\rm m, 0}$ &  $\sigma_8$ \\
\hline
\LCDM & $-$ & $67.69^{+0.43}_{-0.44}$ & $0.3110^{+0.0058}_{-0.0059}$ & $0.811^{+0.007}_{-0.008}$ \\
LV & $0.051^{+0.089}_{-0.087}$ &  $68.05^{+0.76}_{-0.75}$& $0.297^{+0.025}_{-0.026}$ & $0.848^{+0.050}_{-0.067}$ \\ 
\hline 
\end{tabular}
\caption{Values of the cosmological parameters in \LCDM and in the interacting linear vacuum model from \cite{Kaeonikhom:2022ahf}, which were obtained using Planck 2018, BAO, RSD and Pantheon data. All the quoted parameters are
dimensionless besides $\left[H_0\right]=$ km\,s$^{-1}$Mpc$^{-1}$.}\label{tab:LV_table}
\end{table*}

From this figure, we can firstly see that the mildly growing vacuum preferred by the given combination of data results in an enhanced $h(a)$ and suppressed $\Omega_{\rm m}(a)$, exactly as expected from our theoretical analysis shown in figure \ref{fig:LV_theory}. This leads to a slight alleviation of the $H_0$ tension, though a larger value of the coupling strength $q$ would be needed to completely resolve it.

With regard to the quantities which probe the perturbations rather than the background, i.e. $\sigma_8(a)$ and $f \sigma_8(a)$, the interpretation is a little more subtle. From table \ref{tab:LV_table}, the LV $\sigma_8$ is larger than that of \LCDM. As we have seen in figure \ref{fig:LV_today}, our theoretical analysis shows that obtaining $\sigma_8(a)<0.80$ in this model is nearly impossible.
Most importantly, for a resolution of the $\sigma_8$ tension to be achieved by such an interacting model with this dataset, we would need to see the value of $\sigma_8$ decrease with respect to \LCDM: the opposite to what we see here.

A comparison between the theoretical and data-driven results for $f \sigma_8(a)$ reveals a contrasting behaviour. Since $f \approx \Omega_{\rm m}^{\gamma}$, in the case of a growing vacuum, this quantity will be suppressed with respect to \LCDM. On the other hand, $\sigma_8$ will barely change for positive values of $q$ relatively close to zero. These two 
effects imply that the joint quantity $f\sigma_8(a)$ will be enhanced for $q>0$ 
if the value of $\sigma_8$ is substantially larger than its value in \LCDM (as seen in figure \ref{fig:LV_data}). However, for values of the LV $\sigma_8$ close to \LCDM, $f \sigma_8(a)$ will be suppressed overall (as seen in figure \ref{fig:LV_theory}).

In summary, we have seen that the interacting linear vacuum model is potentially capable of alleviating the $H_0$ tension when $q>0$. Unfortunately, this will also result in a worsening of the $\sigma_8$ tension. Furthermore, observational data favours a value of the coupling strength $q$ which does not allow for a significant reduction in the $H_0$ tension.

\subsection{Interacting generalised Chaplygin gas} \label{sec:gCg}
We now examine the interacting generalised Chaplygin gas, characterised by the energy transfer $Q= 3\alpha H (V \rho_{\rm c})/(V + \rho_{\rm c})$.

\subsubsection{Theory plots}\label{sec:gCgtheory}

\begin{figure*} 
\centering 
\includegraphics[width=0.49\textwidth]{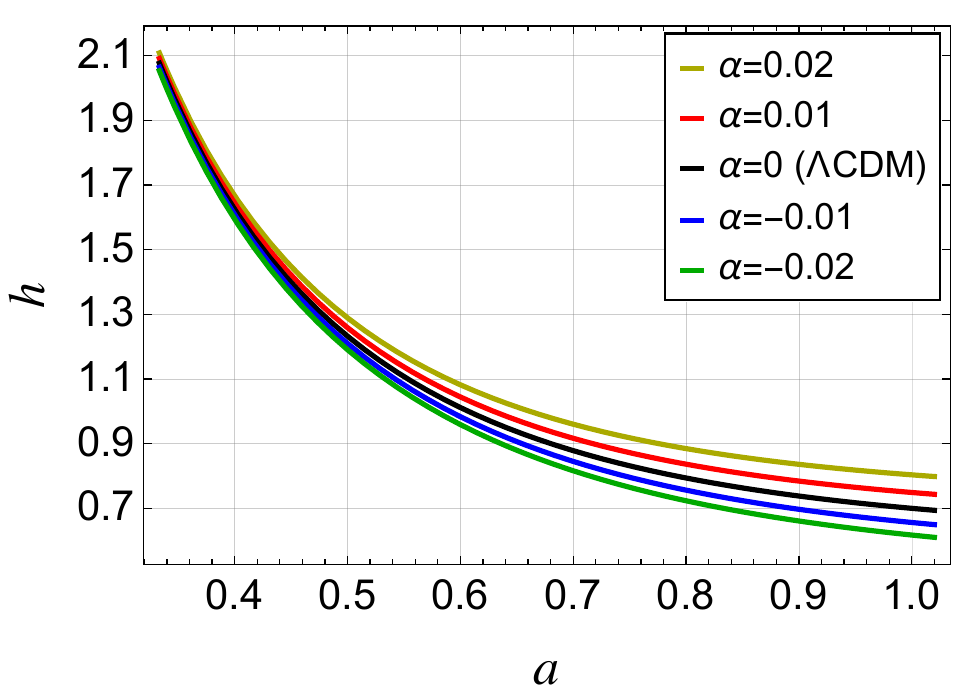}
\includegraphics[width=0.49\textwidth]{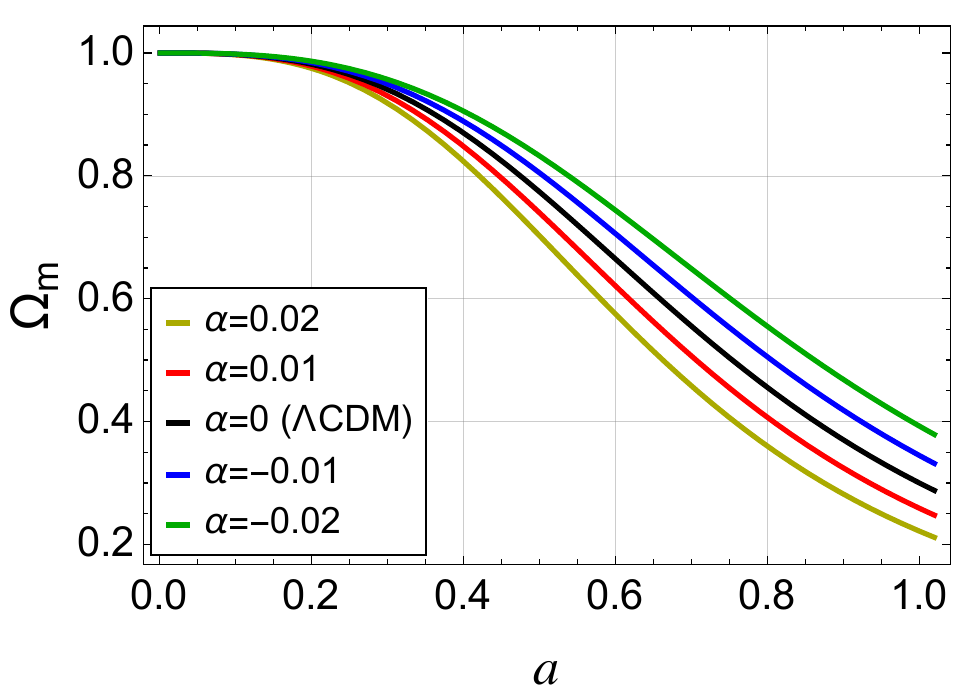}\\
\includegraphics[width=.49\textwidth]{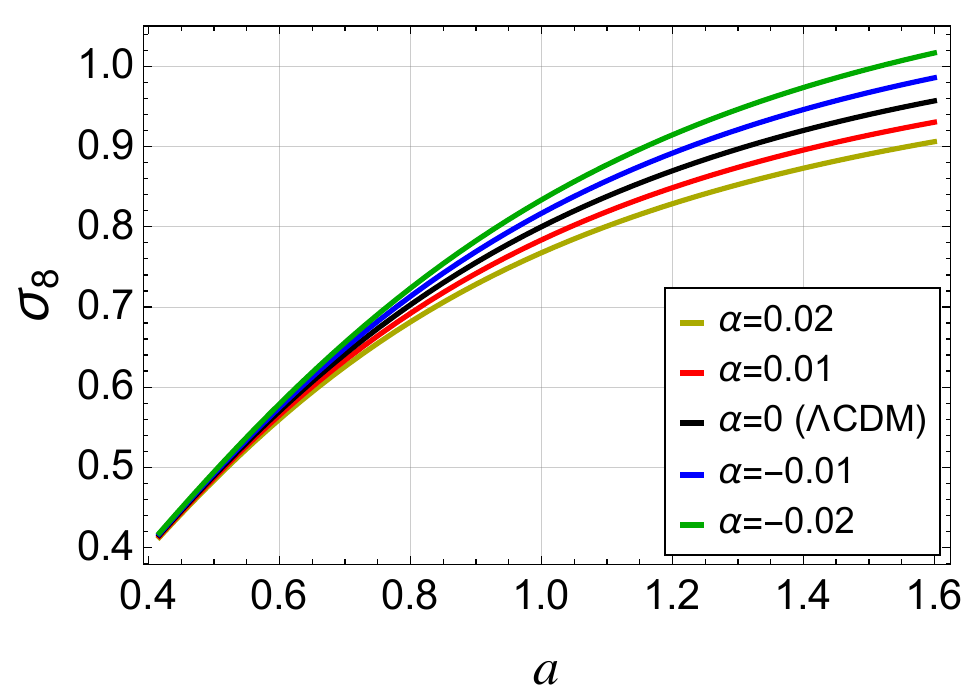}
\includegraphics[width=.49\textwidth]{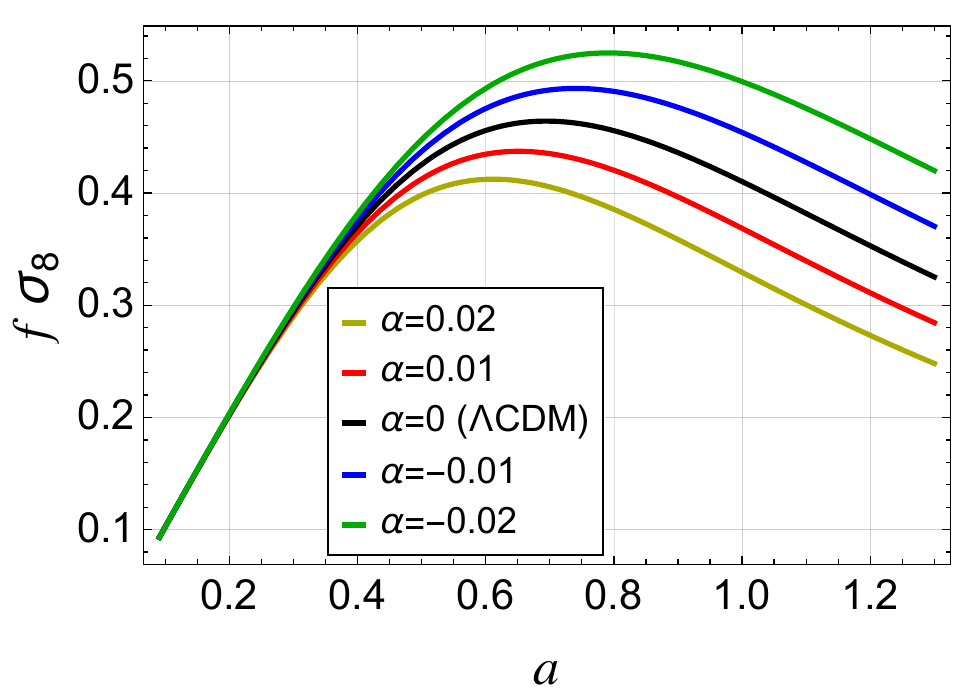}
\caption{Clockwise from top left: $h(a)$, $\Omega_{\rm}(a)$, $f \sigma_8(a)$ and $\sigma_8(a)$ in the interacting generalised Chaplygin gas model. Five different values of the coupling strength $\alpha$ are shown: $\alpha=0.02$ (yellow), $\alpha=0.01$ (red), $\alpha=0$ (black), $\alpha=-0.01$ (blue), $\alpha=-0.02$ (green).}
\label{fig:gCg_theory}
\end{figure*}

In figure \ref{fig:gCg_theory}, we show the evolution of $h(a)$, $\Omega_{\rm m}(a)$, $\sigma_8(a)$ and $f\sigma_{8}(a)$, once again using example values of the coupling strength $\alpha$ between $-0.02$ and $0.02$. The \LCDM cosmology (black), characterised by $h_0=0.70$, $\Omega_{\rm b, 0}=0.05$, $\Omega_{\rm c, 0}=0.25$ and $\sigma_8=0.80$, is recovered for $\alpha=0$.

We can see from the evolution of the quantities shown in this figure that the background behaviour of the interacting generalised Chaplygin gas cosmologies are very similar to that of the linear vacuum cosmologies presented in the previous section. Specifically, a growing vacuum, i.e. $\alpha >0$, yields a smaller $\Omega_{\rm m}(a)$ and a larger $h(a)$ than the \LCDM case.

Differences between the interacting linear vacuum and generalised Chaplygin gas cosmologies become apparent at the level of the first order matter perturbations. Specifically, there is a very apparent difference in the evolution of $\sigma_8(a)$: for a non-zero value of the coupling strength, the curves never cross the \LCDM solution, unlike in the linear vacuum case. For example, the $\sigma_8(a)$ yellow curve, corresponding to $\alpha=0.02$, always remains below the \LCDM case. The same behaviour is reflected in $f\sigma_8(a)$.

\begin{figure*} 
\centering
\includegraphics[width=0.49\textwidth]{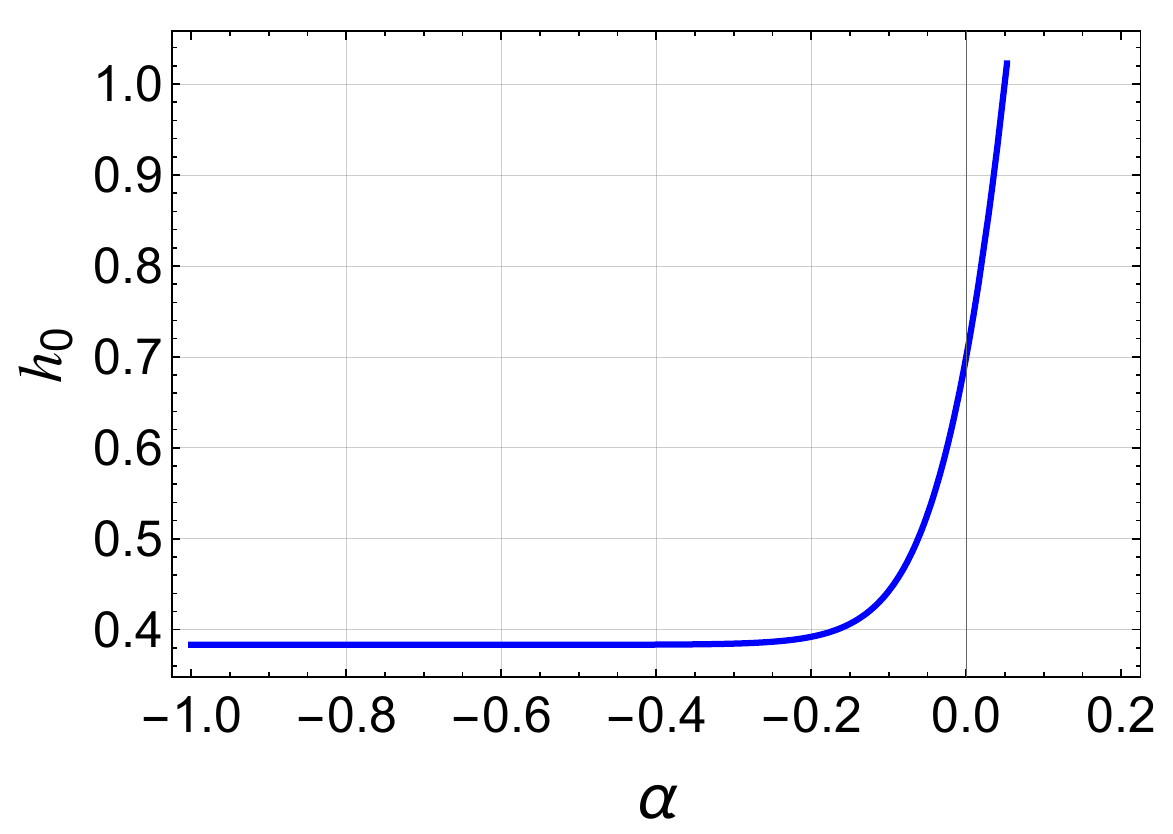}
\includegraphics[width=0.49\textwidth]{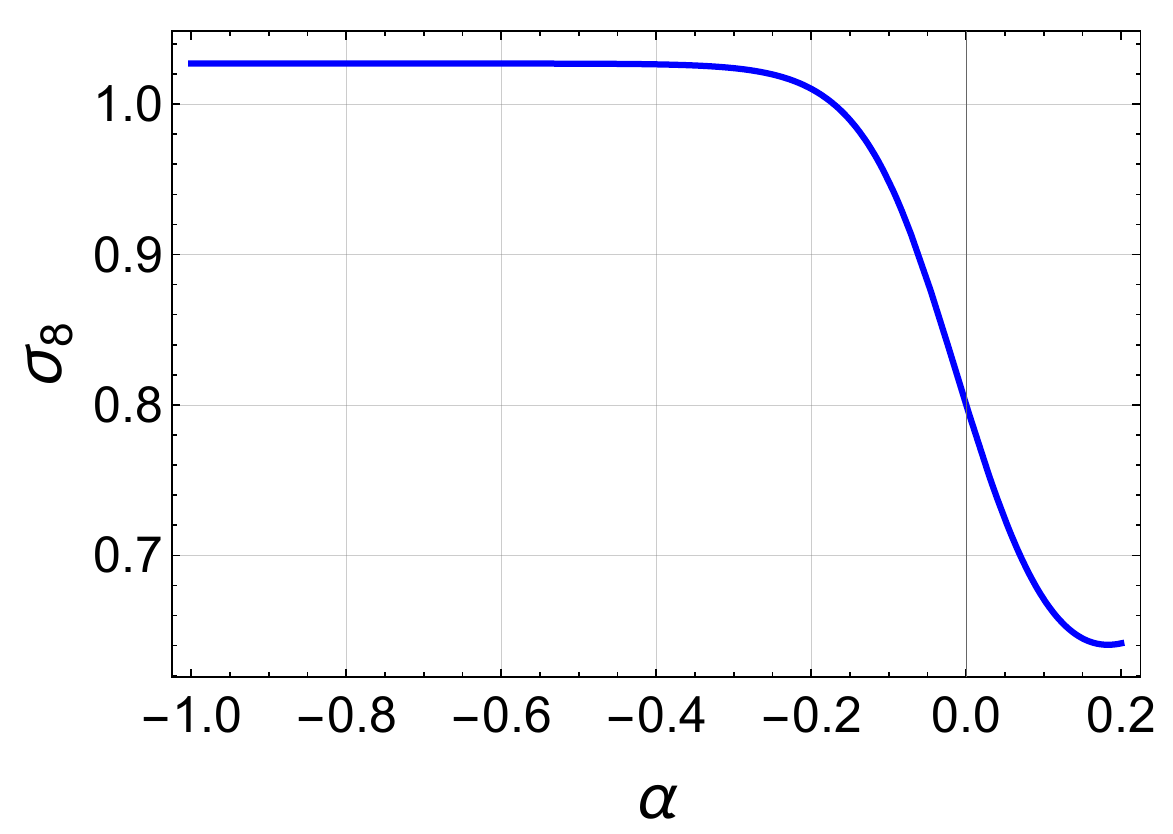}
\caption{\textbf{Left:} how $h_0$ changes as a function of $\alpha$. \textbf{Right:} how $\sigma_8\equiv\sigma_8(z=0)$ changes as a function of $\alpha$. The vertical line in each panel at $\alpha=0$ indicates the \LCDM limit.}
\label{fig:gCg_today}
\end{figure*}

The reason for this difference can be seen in figure \ref{fig:gCg_today}, where we plot $h_0$ and $\sigma_8$ as a function of the coupling strength $\alpha$. The behaviour of $h_0(\alpha)$ is very similar to that of the linear vacuum case, with positive values of the coupling boosting the present time expansion rate. However, unlike in the linear vacuum case, $\sigma_8(\alpha)$ does not display a minimum around the \LCDM limit, $\alpha=0$. Instead, small positive values of $\alpha$ act to decrease $\sigma_8$. This means that both the $H_0$ and $\sigma_8$ tensions have the potential to be resolved in the interacting generalised Chaplygin gas model. Furthermore, this model has only one additional parameter with respect to \LCDM. In the next section, we will examine the observational constraints on this model.

\subsubsection{Data plots}
\begin{figure*} 
\centering 
\includegraphics[width=0.49\textwidth]{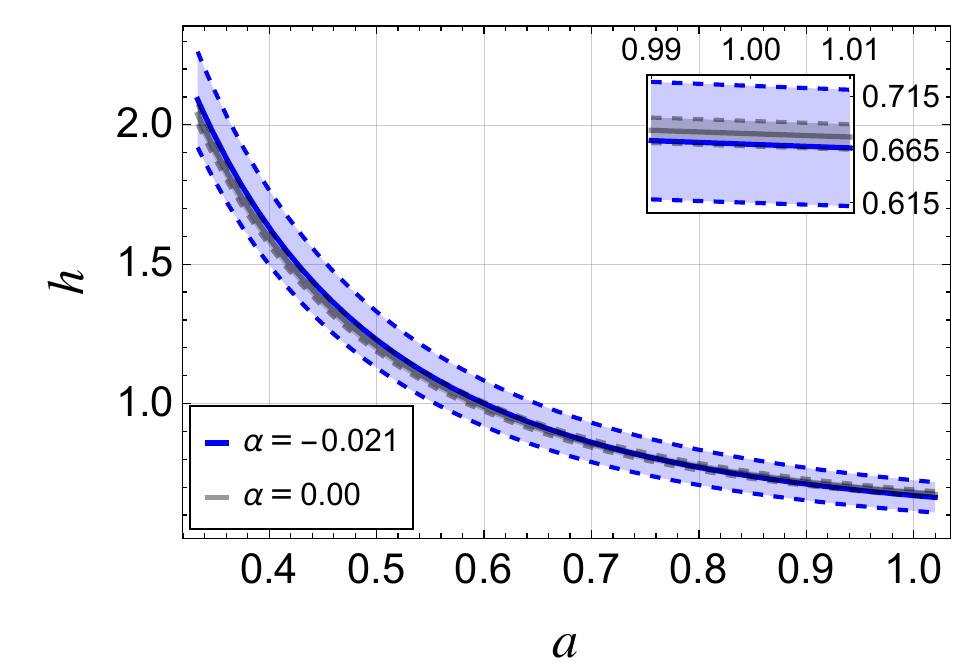}
\includegraphics[width=0.49\textwidth]{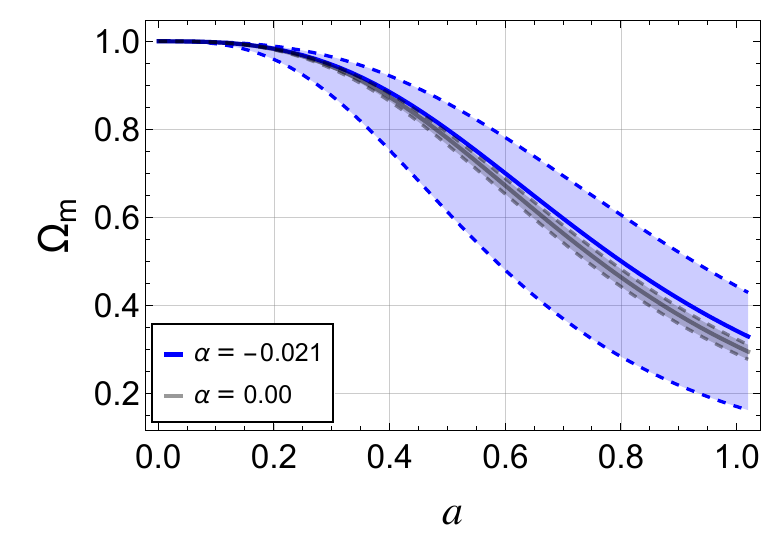}
\includegraphics[width=0.49\textwidth]{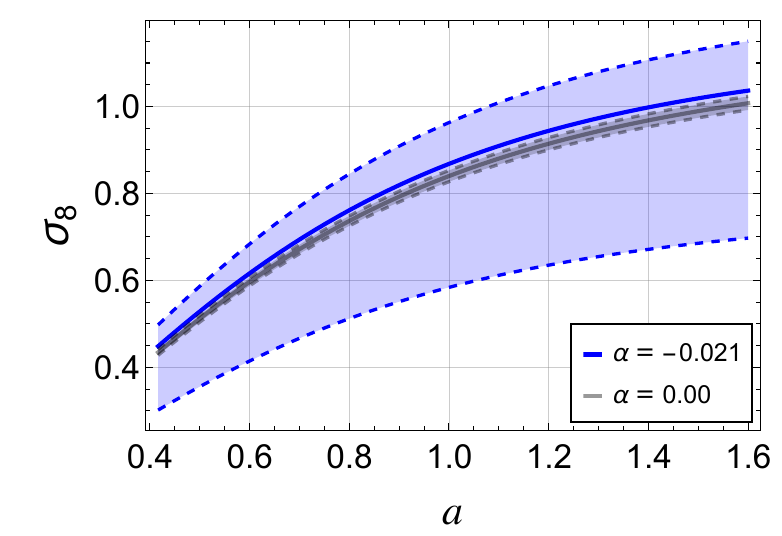}
\includegraphics[width=0.49\textwidth]{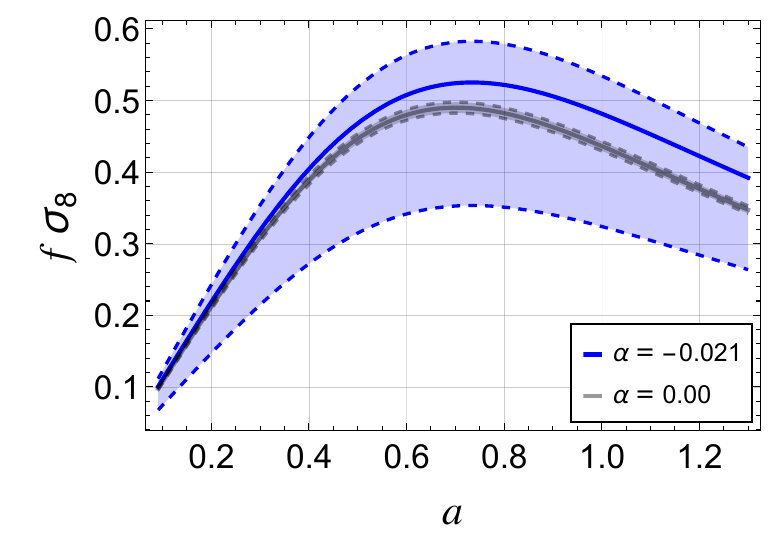}
\caption{Clockwise from top left: $h(a)$, $\Omega_{\rm}(a)$, $f \sigma_8(a)$ and $\sigma_8(a)$ in the interacting generalised Chaplygin gas model as constrained by data. The grey lines show the \LCDM results and the blue lines show the interacting model results as obtained in \cite{Wang:2014xca}. The shaded regions represent the $1\sigma$ error bars.}
\label{fig:gcg_data}
\end{figure*}

As in the case of the linear vacuum, we now plot the evolution of the same cosmological quantities but with their initial conditions provided by observational data. Specifically, we use the results of \cite{Wang:2014xca}, obtained using a pure CMB dataset of Planck 2013 TT \citep{Planck:2013pxb} and WMAP 9 low-$\ell$ polarisation data \citep{Hinshaw2013}. The numerical values are reported in table~\ref{tab:gcg}. The results of this analysis are shown in figure~\ref{fig:gcg_data}.

\begin{table*}
\centering 
\begin{tabular}{Sccccc} 
\hline
\hline 
Model & $\alpha$  & $H_0$ & $\Omega_{\rm m, 0}$ & $\sigma_8$ \\ 
\hline
\LCDM  & $-$ & $68.0^{+1.2}_{-1.2}$ & $0.307^{+0.016}_{-0.018}$ & $0.840^{+0.013}_{-0.013}$ \\
\text{gCg} & $-0.021^{+0.215}_{-0.294}$ & $67.0^{+5.5}_{-5.5}$ & $0.342^{+0.101}_{-0.172}$ & $0.868^{+0.095}_{-0.284}$ \\
\hline 
\end{tabular}
\caption{Values of the cosmological parameters in \LCDM and the interacting generalised Chaplygin gas model from \cite{Wang:2014xca}, which were obtained using Planck 2013 TT and WMAP 9 low-$\ell$ polarisation data. All the quoted parameters are
dimensionless besides $\left[H_0\right]=$ km\,s$^{-1}$Mpc$^{-1}$.}
\label{tab:gcg}
\end{table*}

These results show us that the observational data used in \cite{Wang:2014xca} produces an interacting generalised Chaplygin gas cosmology with a very similar background to \LCDM and a slightly more distinct evolution of the perturbations (as seen in $f\sigma_8(a)$). The data yields a negative value of the coupling strength $\alpha$, which corresponds to a decaying vacuum. As we have seen in figure \ref{fig:gCg_today}, negative coupling strengths lead to smaller values of $H_0$ than in \LCDM, thus worsening the tension in that parameter. We also see the corresponding enhancement in $\sigma_8$ that a decaying vacuum produces in this case. This means that, with this particular combination of data, both cosmological tensions are worsened. We postulate that a re-analysis of the same model with the latest CMB data and with the addition of late-time probes such as BAO and SNIa will likely lead to the interacting model being constrained to be even closer to \LCDM. It is likely the interacting gCg model will never be able to realise its theoretical potential to resolve both tensions.

\subsection{Interacting Shan--Chen}
\label{sec:SC_results}

We now present the results for the interacting Shan--Chen model, in which the coupling function is given by
\begin{equation}
    Q=3qH\left[(1+\beta)V+\frac{\beta G}{2}\rho_*\left(1-\mathrm{e}^{-\alpha\frac{V}{\rho_*}}\right)^2\right] \nonumber.
\end{equation}
In our analysis we fix $\alpha=2.7$ and $G=-8.0$, leaving us with freedom to explore the $q$--$\beta$ parameter space, as done in \cite{Hogg:2021yiz}.

\subsubsection{Theory plots}
\begin{figure*}
\centering 
\includegraphics[width=0.49\textwidth]{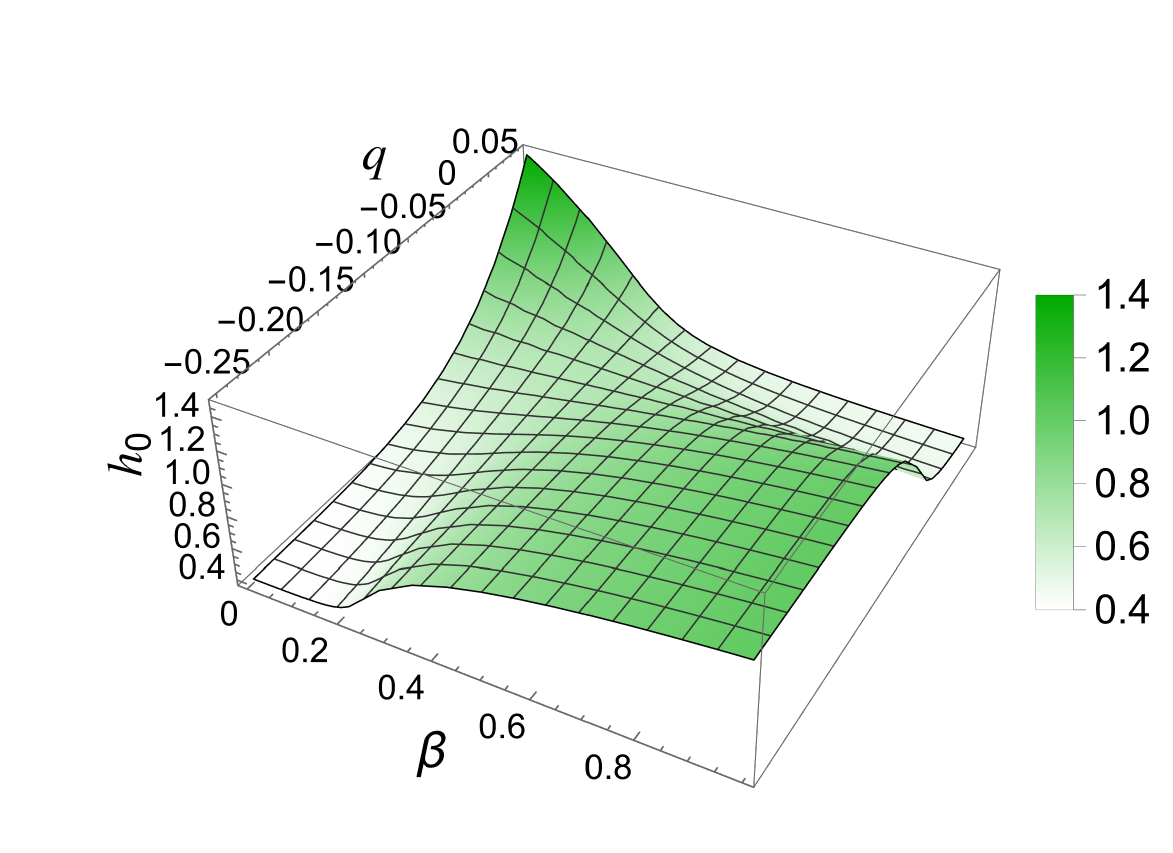}
\includegraphics[width=0.49\textwidth]{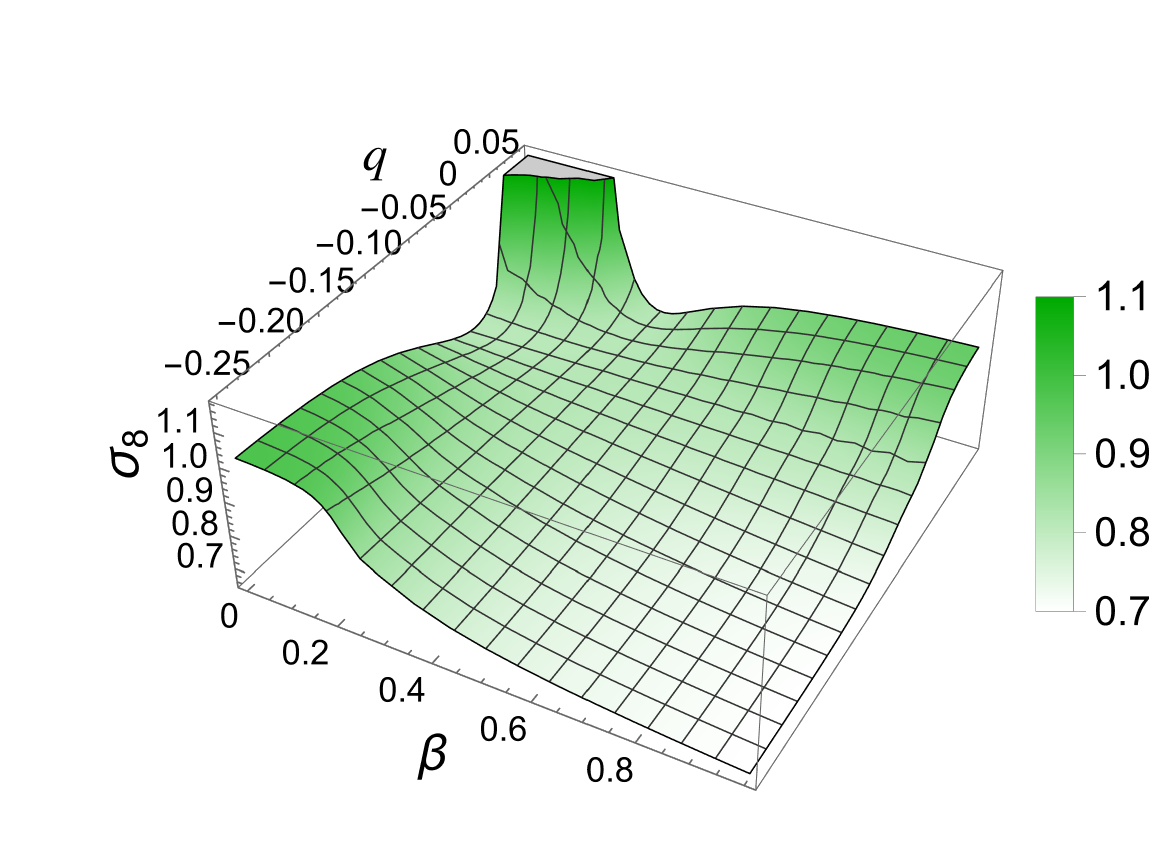}
\caption{\textbf{Left:} how $h_0$ changes as a function of $q$ and $\beta$ in the interacting Shan--Chen model. \textbf{Right:} how $\sigma_8 \equiv \sigma_8(z=0)$ changes as a function of $q$ and $\beta$ in the interacting Shan--Chen model.}
\label{fig:SC_theory}
\end{figure*}

In figure \ref{fig:SC_theory}, we show three-dimensional plots of $h_0$ and $\sigma_8$ as functions of the SC model parameters $q$ and $\beta$. Firstly, it is important to note that the \LCDM limit in this model occurs only when $q=0$; if $q$ is non-zero and $\beta=0$, the interaction is still present, and acts as the LV model discussed above; as for the previous models we fix $h_0=0.70$, $\Omega_{\rm b, 0}=0.05$, $\Omega_{\rm c, 0}=0.25$ and $\sigma_8=0.80$.

From the left-hand panel of this figure, we can see that for positive values of $q$ and small positive values of $\beta$, $h_0$ is enhanced. In addition, $h_0$ is strongly enhanced also for large negative values of $q$ and large positive values of $\beta$.\footnote{While $\beta$ can be both positive or negative, we only consider a positive range for this parameter, since negative values result in numerical issues in the \texttt{Mathematica} computation. Furthermore, \cite{Hogg:2021yiz} found that observational data constrains $\beta > 0$ at around the $2\sigma$ level.} Both of these regions of parameter space correspond to a growing vacuum. This means that there are two separate intervals of $q$ and $\beta$ values which could allow for a resolution of the $H_0$ tension. However, the bimodal behaviour of the $h_0(q, \beta)$ function means that exploration of the parameter space with traditional Metropolis--Hastings MCMC methods may be challenging, if both areas are equally favoured by observational data. This was already seen in the analysis of \cite{Hogg:2021yiz}.

A similarly complex behaviour can be seen in the right-hand panel of figure \ref{fig:SC_theory}, where we plot $\sigma_8$ as a function of the model parameters $q$ and $\beta$. Recalling that $\sigma_8$ ought to be smaller than that of \LCDM for a reduction in tension, we can see that this occurs when $\beta$ is large and positive and when $q$ is large and negative. This is the same region of parameter space which provides a very strong enhancement of $h_0$ -- likely strong enough that $H_0$ would ``overshoot'' the target value of around $73$ km\,s$^{-1}$Mpc$^{-1}$ needed to resolve the tension, creating a new tension in which the cosmology-predicted value from the CMB would be significantly larger than that measured by Type Ia supernovae.

\subsubsection{Data plots}
\begin{figure*}
\centering 
\includegraphics[width=.49\textwidth]{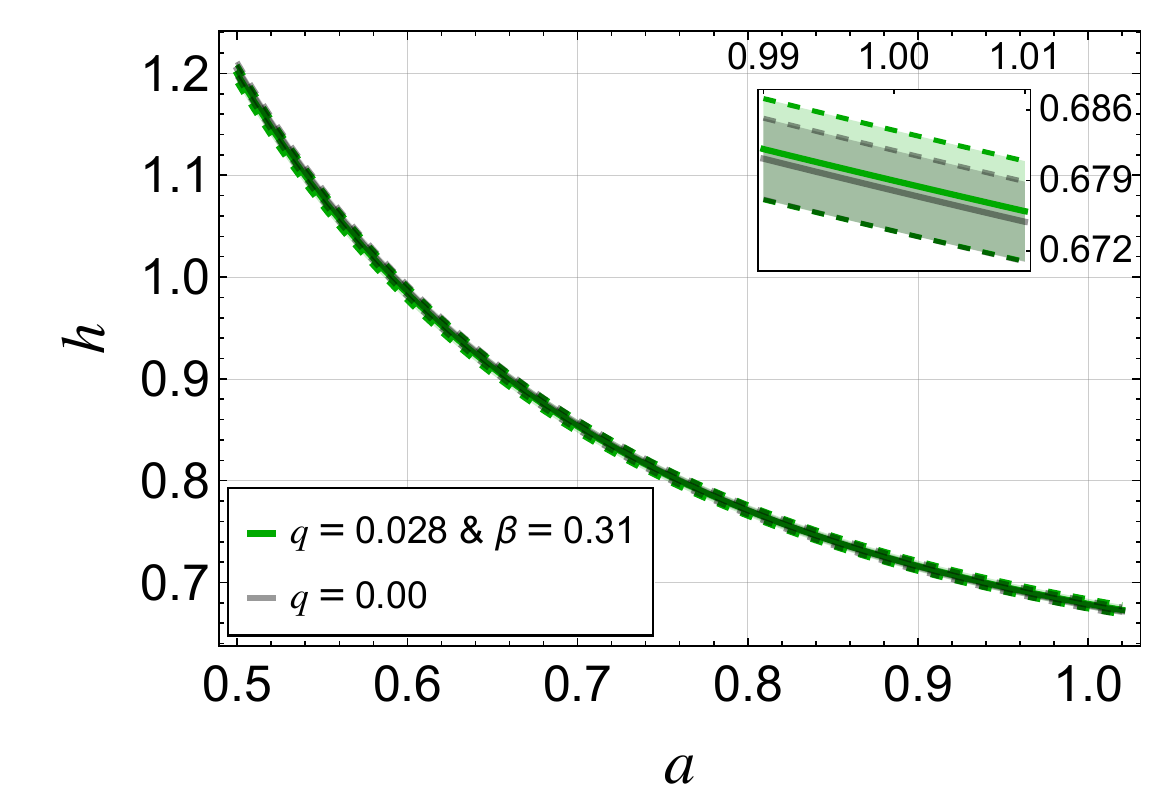}
\includegraphics[width=.49\textwidth]{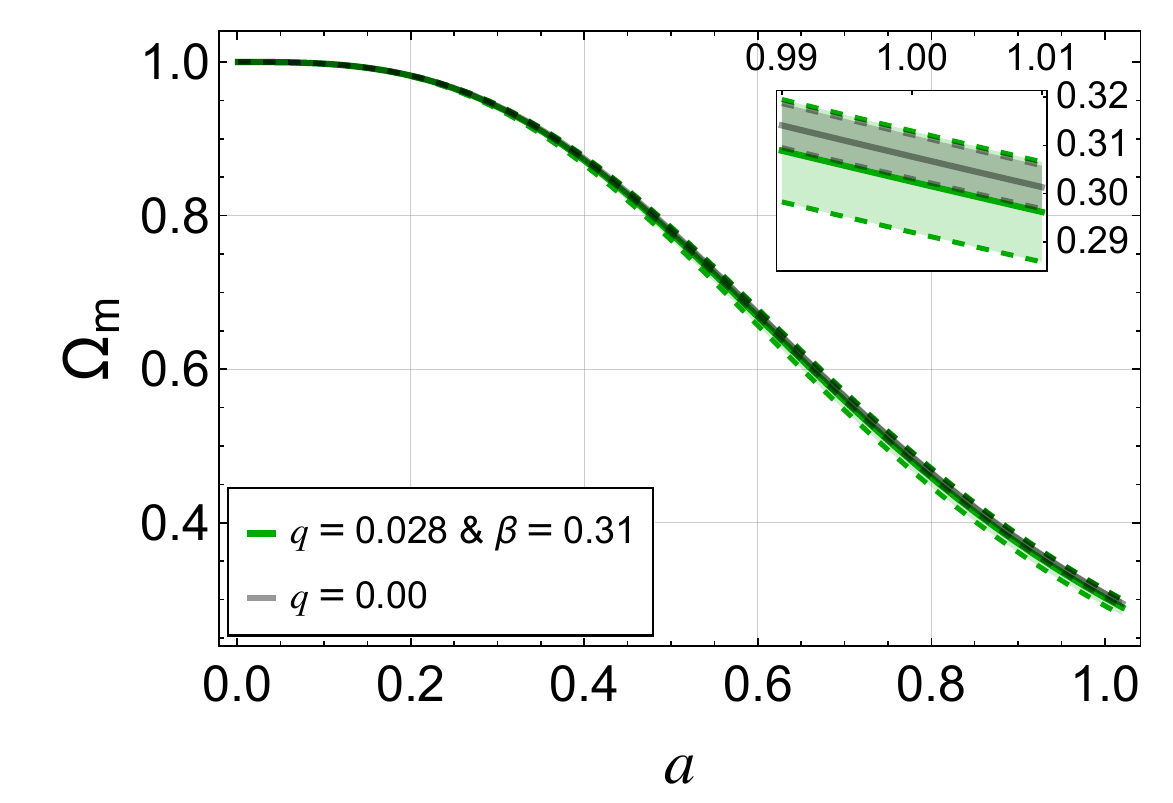}
\includegraphics[width=.49\textwidth]{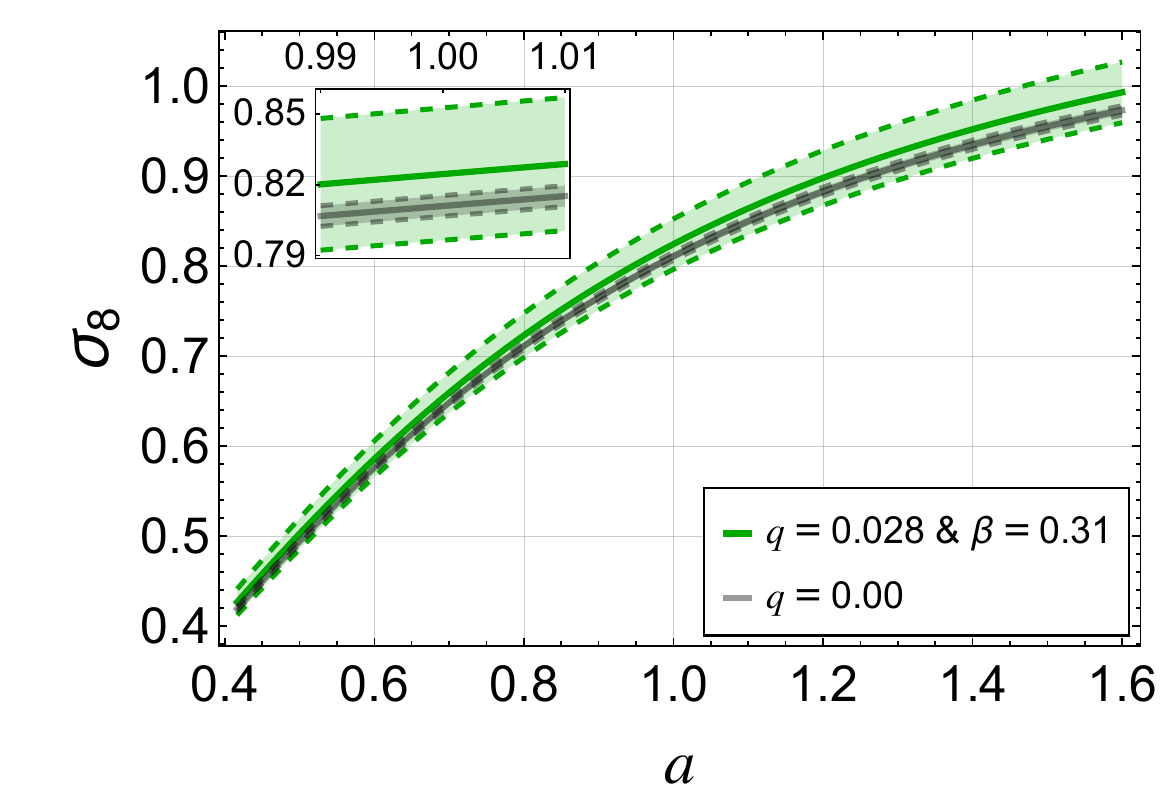}
\includegraphics[width=.49\textwidth]{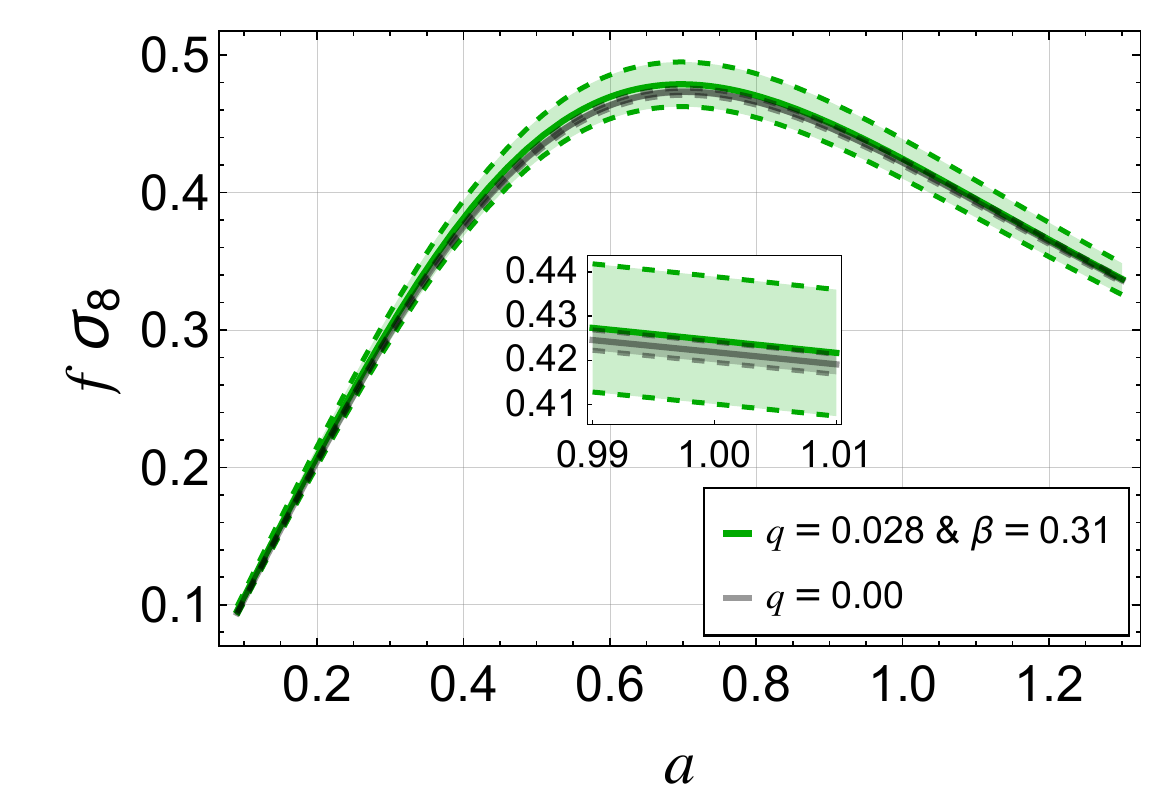}
\caption{Clockwise from top left: $h(a)$, $\Omega_{\rm}(a)$, $f \sigma_8(a)$ and $\sigma_8(a)$ in the interacting Shan--Chen model as constrained by data. The grey lines show the \LCDM results and the green lines show the interacting model results as obtained in \cite{Hogg:2021yiz}. The shaded regions represent the $1\sigma$ error bars.}
\label{fig:SC_data}
\end{figure*}

As in the models studied above, we now explore the evolution of the same cosmological quantities of interest but with their initial conditions informed by results obtained from observational data rather than arbitrary values. In detail, we use the results of \cite{Hogg:2021yiz}, which were obtained using the Planck 2018 TTTEEE CMB measurements \citep{Aghanim:2018eyx}, 6dF and SDSS BAO data \citep{Beutler2011, Ross:2014qpa, Alam:2016hwk} and the Pantheon Type Ia supernova catalogue \citep{Scolnic:2017caz}. The numerical values we use are shown in table~\ref{tab:SC} and the results of this analysis are plotted in figure~\ref{fig:SC_data}.

\begin{table*}
\centering 
\begin{tabular}{Scccccc} 
\hline
\hline 
Model & $q$ & $\beta$ & $H_0$ & $\Omega_{\rm m, 0}$ & $\sigma_8$ \\ 
\hline
\LCDM  & $-$ & $-$ & $67.8\pm0.4$ & $0.3077\pm0.0039$ & $0.8110\pm0.0044$ \\
\text{SC} & $0.028\pm 0.100$ & $0.31\pm0.16$ & $67.9\pm0.5$ & $0.3024\pm0.0097$ & $0.8245\pm0.0280$ \\
\hline 
\end{tabular}
\caption{Values of the cosmological parameters in \LCDM and the interacting Shan--Chen model from \cite{Hogg:2021yiz}, which were obtained using a combination of Planck 2018 CMB TTTEEE data, BAO measurements and the Pantheon SNIa catalogue. All the quoted parameters are
dimensionless besides $\left[H_0\right]=$ km\,s$^{-1}$Mpc$^{-1}$.}
\label{tab:SC}
\end{table*}

From this figure, we can see that the observational data used in \cite{Hogg:2021yiz} constrains the Shan--Chen cosmology to be very close to that of \LCDM, both at the background level and at the level of the first order matter perturbations. The evolution of $h(a)$ is enhanced by an extremely small amount, reflecting the model's potential to relax the tension in the $H_0$ parameter. However, this is accompanied by an enhancement in $\sigma_8(a)$; in other words, the data does not favour the region of parameter space identified in figure \ref{fig:SC_theory} in which both tensions are simultaneously resolved. In fact, the cosmology preferred by the data sits almost exactly at the saddle point in the $h_0(q, \beta)$ and $\sigma_8(q, \beta)$ functions. This may be a result of the sampling method used to obtain the constraints which is not suited to explore multimodal parameter spaces.

\section{Conclusions}
\label{sec:Conlusions}
Our work had two distinct aims: on the one hand, to provide a complete description of the interacting vacuum scenario using the covariant gauge-invariant approach;  and on the other, to explore three specific models of interaction with the goal of identifying why these models may or may not be able to resolve the well-known cosmological tensions. 

In section \ref{sec:IV}, we began by developing the covariant theory of an interacting vacuum, before introducing the gauge-invariant approach and the relevant covariant and gauge-invariant quantities. We provided a physical interpretation of these quantities, and, specialising to a spatially flat FLRW background, imposed the geodesic condition for the CDM particles by mandating that the interaction between CDM and vacuum is a pure energy exchange. In the last part of the section, we developed the evolution of first order perturbations in the CGI approach and, after a brief aside on a specific choice of energy exchange in the geodesic CDM framework, we connected them with
structure growth in cosmology and the familiar observable quantity $\sigma_8$. In section \ref{sec:method} we then introduced the specific models that were studied in this work: the interacting linear vacuum model \citep{Wands:2012vg, Martinelli:2019dau}, the interacting generalised Chaplygin gas model \citep{Wang:2013qy, Wang:2014xca} and the interacting Shan--Chen model \citep{Hogg:2021yiz}.

In section \ref{sec:results} we presented the results from the analysis of the above-described models. In the case of the interacting linear vacuum, we found that, due to a minimum in the function $\sigma_8(q)$ around $q=0$, the two cosmological tensions can never be simultaneously resolved. In fact, for values of the coupling strength sufficiently close to zero, $\sigma_8$ will always remain too close to \LCDM in this model, 
regardless of the direction of the coupling (CDM to vacuum or vice versa). The sign of the coupling does alter the effect of the interaction on the background expansion, with a positive coupling (growing vacuum) leading to an increase in $H_0$.

The interacting generalised Chaplygin gas model had a similar behaviour at the background level, but there is no minimum in the $\sigma_8(\alpha)$ function close to the \LCDM limit of $\alpha=0$. This means that this model has the potential to resolve both tensions simultaneously. This would require a positive value for the coupling strength i.e. a growing vacuum. However, when confronted with observational data, the opposite occurs: both tensions are simultaneously worsened, and a decaying vacuum is favoured. This result was obtained with relatively minimal data: just CMB TT and low-$\ell$ polarisation data. The interacting gCg model was also recently analysed with a larger dataset consisting of Planck 2018 CMB TTTEEE spectra, RSD and JLA SNIa~\citep{Borges:2023xwx}. In that case, a coupling between dark matter and vacuum was indicated by the data, but the cosmological tensions were still not fully resolved.

Lastly, we examined the interacting Shan--Chen model. In this case, we found that there is also a region of parameter space in which both tensions can be resolved simultaneously, but that observational data constrains the model parameters to be close to the \LCDM limit, ending up with a worsening of the $\sigma_8$ tension. Furthermore, we saw that, due to the two additional parameters in this model with respect to \LCDM, the parameter space contains complex bimodal features, making it challenging for a Metropolis--Hastings MCMC analysis to sample from. It would then be interesting for such models to be analysed with, for example, a nested sampling method, to ensure that all the features of the parameter space are fully explored.

From these results we conclude that simple interacting vacuum dark energy models like the linear vacuum model analysed here are not able to resolve the $H_0$ and $\sigma_8$ tensions simultaneously; or one tension may be slightly relaxed, always at the expense of exacerbating the other. In more complicated models like the interacting Chaplygin gas and the interacting Shan--Chen model, there exist regions of parameter space where both tensions may simultaneously be resolved, but observational data tends to constrain the model parameters to be extremely close to their \LCDM limits, irrespective of the existence of such regions. Since all of these models have additional parameters with respect to \LCDM, they will likely be disfavoured from a statistical standpoint. This can be seen in the $\chi^2$ values for the Shan--Chen model presented in \cite{Hogg:2021yiz} (all larger than \LCDM) and the Bayes factors for a time-dependent linear vacuum found by \citep{Hogg:2020rdp} (no evidence that the model should be preferred over \LCDM). 

Our results are in agreement with recent works which studied dark energy models which have small deviations from a \LCDM background expansion, or small deviations from Newton's constant represented by an effective gravitational constant, $G_{\rm eff}$ \citep{Heisenberg:2022gqk, Heisenberg:2022lob}. In these cases, it was shown that a simultaneous resolution of the $H_0$ and $\sigma_8$ tensions is very difficult, with the evolving dark energy equation of state $w(z) = p/\rho$ being required to cross the phantom divide i.e. $w=-1$ at some point in its history, or else $G_{\rm eff}$ needing to be significantly reduced. For the latter criterion, interacting models with momentum transfer may provide a way forward \citep{Pourtsidou:2016ico, Linton:2021cgd}.

In conclusion, if statistically convincing resolutions to both cosmological tensions are simultaneously sought, other avenues besides the simple interacting vacuum scenarios which only have energy exchange studied here should be prioritised for exploration. However, in light of the recent reduction of the $\sigma_8$ tension with the \texttt{Hillipop} Planck likelihood \citep{Tristram:2023haj}, it may only be necessary to search for models in which the $H_0$ tension alone is resolved. Furthermore, exploring models with spatial curvature may be a promising avenue for investigation \citep{Yang:2022kho}.


\section*{Acknowledgements}

We are grateful to Eoin Ó Colgáin for useful comments. MS is supported by a doctoral studentship of the Science and Technology Facilities Council (STFC): Training Grant No. ST/X508822/1, Project Ref. 2753640. NBH is supported by a postdoctoral position funded by IN2P3. MB was supported by UK STFC Grant No. ST/S000550/1 and  ST/W001225/1.

\section*{CRedIT statement} 
\textbf{Marco Sebastianutti}:~Conceptualisation; Methodology; Software; Visualisation; Formal analysis; Writing -- original draft; Writing -- review \& editing. \textbf{Natalie B. Hogg}:~Conceptualisation; Validation; Visualisation; Writing -- original draft; Writing -- review \& editing. \textbf{Marco Bruni}:~Conceptualisation; Methodology; Visualisation; Supervision; Writing -- review \& editing.

\appendix
\section{Equivalence of the CGI and Bardeen variables at first order}\label{app:A}

The standard way to deal with perturbations in cosmology is to use the Bardeen variables \citep{Bardeen:1980kt}. These are linear combinations of gauge-dependent quantities built such that they are gauge-invariant at first order in perturbations.
The Bardeen approach, unlike the CGI one, is based on coordinates; therefore its variables acquire a physical meaning only once a specific gauge is chosen.

By expanding the CGI quantities into a background and a first order part, we recover the corresponding gauge-invariant Bardeen variables. In this appendix we show the equivalence, \textit{at first order}, of the two methods by decomposing a particularly relevant CGI quantity, the comoving fractional density spatial gradient $\Delta^\mu$. We refer the reader to \cite{Bruni:1992dg} for a complete list of identities.

We start from the linearly perturbed spatially flat FLRW metric in the form:
\begin{align}
    {\rm d}s^2&=-(1+2\phi){\rm d}t^2+2a\partial_iB\,{\rm d}t\,{\rm d}x^i \nonumber \\
    &+a^2[(1-2\psi)\delta_{ij}+2\partial_i\partial_jE]{\rm d}x^i{\rm d}x^j,
\end{align}
where we take into account scalar perturbations only.
In such a spacetime, an exact variable, for instance a scalar $f$, gets split into a homogeneous background part, denoted with a bar, and an inhomogeneous linear perturbation, i.e. $f(t,x^i)=\bar{f}(t) + \delta f(t, x^i)$. In general, this splitting is not a covariant procedure and leads to a gauge (or coordinate) dependence as can be noticed by looking at \eqref{ftimegradient}.
The perturbed 4-velocity of a perfect fluid, $u^\mu=\bar{u}^\mu+\delta u^\mu$, is given by:
\begin{align}
     u_\mu=[-1-\phi,\partial_i\theta], \quad\quad u^\mu=[1-\phi,a^{-1}\partial^i v],
\end{align}
where $\theta=a(v+B)$.

The time derivative of the scalar $f$ reads:
\begin{align}\label{ftimegradient}
    \dot{f}&\equiv u^\mu\nabla_\mu f,\notag\\
    &=\left(\bar{u}^\mu+\delta u^\mu\right)\nabla_\mu\left(\bar{f}+\delta f\right),\notag\\
    &=\dot{\bar{f}}+\left(\delta f\right)^\cdot-\phi\dot{\bar{f}}.
\end{align}
Projecting with the spatial metric $h_{\mu\nu}$ at linear order
\begin{align}\label{fspacegradient}
    h_{\mu}^{\,\,\,\nu}&=\bar{h}_\mu^{\,\,\,\nu}+\delta h_{\mu}^{\nu},\\
    &=\bar{h}_\mu^{\,\,\,\nu}+\bar{u}_\mu\delta u^\nu+\delta u_\mu\bar{u}^\nu,
\end{align}
we obtain the spatial gradient of $f$:
\begin{align}
    \leftindex^{(3)}\nabla_\mu f&\equiv h_\mu^{\,\,\,\nu}\nabla_\nu f,\notag\\
    &=\bar{h}_\mu^{\,\,\,\nu}\nabla_\nu\left(\bar{f}+\delta f\right)+\bar{u}_\mu\delta u^\nu \nabla_\nu\bar{f}+\delta u_\mu\bar{u}^\nu\nabla_\nu\bar{f},\notag\\
    &=\leftindex^{(3)}\nabla_\mu\delta f+\bar{u}_\mu\delta u^\nu\nabla_\nu\bar{f}+\delta u_\mu\dot{\bar{f}},\notag\\
    &=\left[-\delta u^0\dot{\bar{f}}+\delta u_0\dot{\bar{f}},\partial_i\delta f+\delta u_i\dot{\bar{f}}\right],\notag\\
    &=\left[0,\partial_i\left(\delta f+\theta\dot{\bar{f}}\right)\right].\label{fexp}
\end{align}
From the above expression, the linearised comoving fractional density spatial gradient is easily derived
\begin{align}
    \Delta_\mu\equiv\frac{a}{\rho}\leftindex^{(3)}\nabla_\mu\,\rho=\left[0,a\,\partial_i\left(\delta +\theta\dot{\bar{\rho}}/\bar{\rho}\right)\right]=\left[0,a\,\partial_i\Delta_{\textcolor{black}{\rm B}} \right],
\end{align}
where $\delta=\delta\rho/\bar{\rho}$ is the gauge-dependent density contrast and $\Delta_{\textcolor{black}{\rm B}}\equiv\delta+\theta\dot{\bar{\rho}}/\bar{\rho}$ the corresponding first order gauge-invariant Bardeen variable. From \eqref{eq:trace} we can expand the trace of $\Delta_{\mu\nu}$,
\begin{align}
    \Delta\equiv a\leftindex^{(3)}\nabla_\mu\Delta^\mu=\left[0,\partial_i \partial^i\left(\delta +\theta\dot{\bar{\rho}}/\bar{\rho}\right)\right]=\left[0,\nabla^2\Delta_{\textcolor{black}{\rm B}} \right]
\end{align}
and obtain an explicitly-comoving expression, where $\nabla^2\equiv\partial_i\partial^i$. This proves the equivalence of the two approaches to cosmological perturbation theory at linear order.


\section{First order ODEs for density and expansion perturbations}\label{app:B}
In this appendix we derive the first order ODEs for $\Delta_\mu$ and $Z_\mu$ in~\eqref{eq:D1} and~\eqref{eq:Z1} respectively. In order to arrive at the final forms shown in the main text, we make use of some useful identities, valid \textit{at first order} only, involving time derivatives and spatial gradients of scalar quantities.

From the spatial gradient of the continuity equation~\eqref{continuity2} with $p=0$ we find
\begin{align}\label{eq:SpatDotRho}
    \leftindex^{(3)}\nabla_\mu\dot{\rho}&=-\rho\leftindex^{(3)}\nabla_\mu\Theta-3H\leftindex^{(3)}\nabla_\mu\,\rho-\leftindex^{(3)}\nabla_\mu Q,\nonumber\\
    &=-\frac{\rho}{a}Z_\mu-\frac{\rho}{a}3H\Delta_\mu-\leftindex^{(3)}\nabla_\mu Q.
\end{align}
To express the LHS in terms of our standard CGI variables we rewrite the spatial gradient of the time derivative of the energy density $\rho$ as
\begin{align}
    \leftindex^{(3)}\nabla_\mu\dot{\rho} =&\, h_{\mu}^{\,\,\,\sigma}\nabla_\sigma\left(u^\nu\nabla_\nu \rho\right), \\
    =&\, h_{\mu}^{\,\,\,\sigma}\left[\left(\nabla_\sigma u^\nu\right)\left(\nabla_\nu\rho\right)+u^\nu\nabla_\sigma\nabla_\nu\rho\right],\\
     =&\, h_{\mu}^{\,\,\,\sigma}\left[\left(\omega_{\sigma}^{\,\,\,\nu}+\sigma_{\sigma}^{\,\,\,\nu}+\frac{1}{3}\Theta h_{\sigma}^{\,\,\,\nu}- u_\sigma a^\nu\right)\left(\nabla_\nu\rho\right) \right. \nonumber \\
    &+ \left. \left(\nabla_\sigma \rho\right)^{\cdot} \vphantom{\frac13} \right],\\
    =&\, H\leftindex^{(3)}\nabla_\mu\rho+h_{\mu}^{\,\,\,\sigma}\left(\nabla_\sigma \rho\right)^{\cdot},
\end{align}

where we have kept everything at first order in perturbations. We can write the second term in the last line as
\begin{align}
    h_{\mu}^{\,\,\,\sigma}\left(\nabla_\sigma \rho\right)^{\cdot}&=\left(\leftindex^{(3)}\nabla_\mu\rho\right)^\cdot-\left(h_{\mu}^{\,\,\,\sigma}\right)^\cdot\left(\nabla_\sigma \rho\right),\\
    &=\left(\leftindex^{(3)}\nabla_\mu\rho\right)^\cdot-\left(a_\mu u^\sigma+a^\sigma u_\mu\right)\nabla_\sigma \rho,\\
    &=\left(\leftindex^{(3)}\nabla_\mu\rho\right)^\cdot-\dot{\rho}\,a_\mu,
\end{align}
and substitute to find
\begin{align}
    \leftindex^{(3)}\nabla_\mu\dot{\rho}&=H\leftindex^{(3)}\nabla_\mu\rho-\dot{\rho}\,a_\mu+\left(\leftindex^{(3)}\nabla_\mu\,\rho\right)^{\cdot}.\label{eq:spatialdotrho}
\end{align}
Notice that~\eqref{eq:V1} is obtained following the same steps as in the above with the vacuum energy $V$ replacing the density $\rho$. From the definition of $\Delta_\mu$ in~\eqref{list} we have
\begin{align}
    \dot{\Delta}_\mu\equiv\left(\frac{a}{\rho}\leftindex^{(3)}\nabla_\mu\, \rho\right)^\cdot=H\Delta_\mu-\frac{\dot{\rho}}{\rho}\Delta_\mu+\frac{a}{\rho}\left(\leftindex^{(3)}\nabla_\mu\,\rho\right)^{\cdot},
\end{align}
and expressing the last term on the RHS as a function of the comoving fractional density spatial gradient, we can substitute it in~\eqref{eq:spatialdotrho} and, via the momentum conservation relation~\eqref{acceleration} which for vanishing pressure reads $a_\mu=\rho^{-1}V_\mu$, find $\leftindex^{(3)}\nabla_\mu\dot{\rho}$ in terms of $\Delta_\mu$ and $V_\mu$:
\begin{align}
    \leftindex^{(3)}\nabla_\mu\,\dot{\rho}&=\frac{\rho}{a}\left(\dot{\Delta}_\mu+\frac{\dot{\rho}}{\rho}\Delta_\mu-\frac{a\dot{\rho}}{\rho^2}V_\mu\right).
\end{align}
Inserting this back into~\eqref{eq:SpatDotRho} leads to the first order ODE~\eqref{eq:D1}.

From the spatial gradient of the Raychaudhuri equation~\eqref{eq:Raychaudhuri} with vanishing pressure we have
\begin{align}\label{eq:SpatDotTheta}
    \leftindex^{(3)}\nabla_\mu\dot{\Theta}&=-\frac{2H}{a}Z_\mu+\leftindex^{(3)}\nabla_\mu A-\frac{\kappa\rho}{2a}\Delta_\mu+\kappa V_\mu.
\end{align}
If, instead of density, we consider the expansion scalar $\Theta$, the analogue of~\eqref{eq:spatialdotrho} is
\begin{align}
    \leftindex^{(3)}\nabla_\mu\dot{\Theta}&=H\leftindex^{(3)}\nabla_\mu\Theta-\left(3H\right)^\cdot a_\mu+\left(\leftindex^{(3)}\nabla_\mu\,\Theta\right)^{\cdot},
\end{align}
which, thanks to the relation
\begin{align}
    \dot{Z}_\mu\equiv\left(a\leftindex^{(3)}\nabla_\mu\Theta\right)^\cdot=HZ_\mu+a\left(\leftindex^{(3)}\nabla_\mu\Theta\right)^{\cdot},
\end{align}
becomes
\begin{align}\label{eq:SpatDotThetaLHS}
    \leftindex^{(3)}\nabla_\mu\dot{\Theta}&=-3\dot{H} a_\mu+\frac{\dot{Z}_\mu}{a}.
\end{align}
Through~\eqref{eq:bgRay} and~\eqref{FriedmannC} and the momentum constraint~\eqref{acceleration} we obtain the below relations at zeroth and first order in perturbations respectively
\begin{align}
    \dot{H}&=-\frac{\kappa}{2}\rho,\\
    A&=\frac{1}{\rho}\leftindex^{(3)}\nabla_\nu \leftindex^{(3)}\nabla^\nu V,
\end{align}
and, substituting these and~\eqref{eq:SpatDotThetaLHS} with $a_\mu=\rho^{-1}V_\mu$ into~\eqref{eq:SpatDotRho}, we arrive at the first order ODE~\eqref{eq:Z1} after swapping $\leftindex^{(3)}\nabla_\mu$ and the spatial Laplacian $\leftindex^{(3)}\nabla^2\equiv\leftindex^{(3)}\nabla_\nu \leftindex^{(3)}\nabla^\nu$ in $\leftindex^{(3)}\nabla_\mu A$.

\section{Covariant gauge-invariant identities at linear order in perturbations}\label{app:C}

In this appendix we present some useful identities, valid \textit{at first order} in perturbations, involving time derivatives and spatial gradients of CGI variables. These relations are employed, starting from~\eqref{2ODE} and using~\eqref{eq:trace} to extract the trace part of the spatial gradient of $\Delta^\mu_{\rm m}$, to arrive at the second order ODE~\eqref{Da}.

For the generic spatial vector $X^\nu$ which vanishes in the background, as for the comoving fractional density spatial gradient $\Delta^\nu$, at first order in perturbations we have:
\begin{align}
    a\leftindex^{(3)}\nabla_\nu\dot{X}_\mu&=a\,h_{\nu}^{\,\,\,\rho}h_{\mu}^{\,\,\,\sigma}\nabla_\rho\left(u^\delta\nabla_\delta X_\sigma\right),\label{eq:Bfirst}\\
    &=a\,h_{\nu}^{\,\,\,\rho}h_{\mu}^{\,\,\,\sigma}\left(H h_{\rho}^{\,\,\,\delta}\nabla_\delta X_\sigma+u^\delta\nabla_\rho\nabla_\delta X_\sigma\right),\label{eq:Bsecond}\\
    &=a\,H\leftindex^{(3)}\nabla_\nu X_\mu+a\left(\leftindex^{(3)}\nabla_\nu X_\mu\right)^\cdot,\label{eq:Bthird}\\
    &=\left(a\leftindex^{(3)}\nabla_\nu X_\mu\right)^\cdot.\label{eq:Bfourth}
\end{align}
To go from the second term in \eqref{eq:Bsecond} to the one in \eqref{eq:Bthird}, we have skipped a couple of intermediate steps: given that the Riemann tensor measures the non-commutativity of covariant derivatives,
\begin{align}
    \left[\nabla_\rho,\nabla_\delta\right]X_\sigma=R_{\tau\sigma\rho\delta}X^\tau
\end{align}
and, exploiting its background form \citep{Ellis:1987zz,Gurses:2020kpv}:
\begin{align}
    R_{\tau\sigma\rho\delta}&=\frac{\kappa}{2}\left(\rho+p\right)\left(u_\tau u_\rho g_{\sigma\delta}+u_\sigma u_\delta g_{\tau\rho}-u_\tau u_\delta g_{\sigma\rho}-u_\sigma u_\rho g_{\tau\delta}\right)\notag\\
    &+\frac{\kappa}{3}\left(\rho+V\right)\left(g_{\tau\rho}g_{\sigma\delta}-g_{\tau\delta}g_{\sigma\rho}\right),
\end{align}
we can commute the covariant derivatives in the second term in~\eqref{eq:Bsecond} and bring the time derivative piece past the two 3-metrics. In fact, after commuting the covariant derivatives, we have that
\begin{align}
    u^\delta h_{\nu}^{\,\,\,\rho}\nabla_\delta\nabla_\rho X_\sigma&=\left(h_{\nu}^{\,\,\,\rho}\nabla_\rho X_\sigma\right)^\cdot-\left(h_{\nu}^{\,\,\,\rho}\right)^\cdot\left(\nabla_\rho X_\sigma\right) \\
    &=\left(h_{\nu}^{\,\,\,\rho}\nabla_\rho X_\sigma\right)^\cdot,
\end{align}
with the time derivative $\left(h_{\nu}^{\,\,\,\rho}\right)^\cdot=2a_{(\nu} u^{\rho)}$ giving rise to terms that are neglected at first order in perturbations.

The same procedure and results are in order for higher time derivatives, e.g. we have
\begin{align}
    a\leftindex^{(3)}\nabla_\nu\ddot{X}_\mu&=\left(a\leftindex^{(3)}\nabla_\nu X_\mu\right)^{\cdot\cdot},
\end{align}
and, as for \eqref{eq:trace}, the trace of $a\leftindex^{(3)}\nabla_\nu\dot{X}_\mu$ and $a\leftindex^{(3)}\nabla_\nu\ddot{X}_\mu$ read:
\begin{align}
    a\leftindex^{(3)}\nabla^\mu\dot{X}_\mu&=\left(a\leftindex^{(3)}\nabla^\mu X_\mu\right)^\cdot\equiv\dot{X},\\
    a\leftindex^{(3)}\nabla^\mu\ddot{X}_\mu&=\left(a\leftindex^{(3)}\nabla^\mu X_\mu\right)^{\cdot\cdot}\equiv\ddot{X}.
\end{align}

\section{Geodesic interaction in the comoving-orthogonal gauge}\label{app:D}
In this brief appendix we aim at showing that, for a geodesic interaction ($f^\mu=0$) between dust
and vacuum, by choosing a frame comoving with the fluid ($v=0$) and requiring orthogonality of the spatial hypersurfaces to its 4-velocity ($B=0$), we automatically find ourselves in a synchronous gauge ($\phi=0$).

Expanding the perfect fluid 4-acceleration $a^\mu$:
\begin{align}\label{eq:acc1}
    a_\mu &= \left(\partial_\nu u_\mu -\Gamma^{\rho}_{\mu \nu }u_\rho\right)u^\nu,\nonumber\\
    &= \bar{a}_\mu+\left(\partial_\nu\bar{u}_\mu \delta u^\nu+\partial_\nu\delta u_\mu \bar{u}^\nu \right. \nonumber \\
    &- \left. \bar{\Gamma}^{\rho}_{\mu \nu}\bar{u}_\rho\delta u^\nu-\bar{\Gamma}^{\rho}_{\mu \nu}\delta u_\rho\bar{u}^\nu-\delta\Gamma^{\rho}_{\mu \nu}\bar{u}_\rho\bar{u}^\nu\right),\nonumber\\
    &=\left[\left(\delta u_0\right)^\cdot+a^{-1}\partial^j v\, \bar{\Gamma}^{0}_{0j}-\partial_j\theta\,\bar{\Gamma}^{j}_{00}+\delta\Gamma^{0}_{00},\left(\delta u_i\right)^\cdot \right. \nonumber \\
    &+ \left. a^{-1}\partial^j v\,\bar{\Gamma}^{0}_{ij}-\partial_j\theta\,\bar{\Gamma}^{j}_{i0}+\delta\Gamma^{0}_{i0}\right],\nonumber\\
    &= \left[0,\partial_i(\dot{\theta}+\phi)\right],
\end{align}

we notice that, in a comoving-orthogonal frame for which $\theta=0$, the geodesic requirement implies $\phi=0$.

\bibliographystyle{elsarticle-num-names}
\bibliography{pdu_submission}

\end{document}